\documentclass[a4paper,aps,prd,showpacs,nofootinbib,preprintnumbers,amsmath,amssymb,mcite,superscriptaddress,12pt]{revtex4-1}

\usepackage[table]{xcolor}
\usepackage{graphicx}
\usepackage{dcolumn}
\usepackage{bm}
\usepackage{esvect}
\usepackage{accents}

\usepackage{hyperref}
\hypersetup{colorlinks=true,linkcolor=magenta,anchorcolor=green,citecolor=cyan,filecolor=black,menucolor=black,urlcolor=brown}
\usepackage{slashed}

\newcommand{\be}{\begin{equation}}
\newcommand{\ee}{\end{equation}}
\newcommand{\ba}{\begin{eqnarray}}
\newcommand{\ea}{\end{eqnarray}}

\begin{document}

~\rightline {IPARCOS-UCM-23-021, CERN-TH-2023-049}

\title{The Invisible Dilaton}

\author{Philippe Brax}
\affiliation{Institut de Physique Th\'eorique, Universit\'e  Paris-Saclay, CEA, CNRS, F-91191 Gif-sur-Yvette Cedex, France}
\affiliation{CERN, Theoretical Physics Department, Geneva, Switzerland.}
\author{Clare Burrage}
\affiliation{University of Nottingham, United Kingdom}
\author{Jose A.R. Cembranos}
\affiliation{Departamento de F\'isica Te\'orica and IPARCOS, Facultad de Ciencias F\'isicas, \\
Universidad Complutense de Madrid, Ciudad Universitaria, 28040 Madrid, Spain}
\author{Patrick Valageas}
\affiliation{Institut de Physique Th\'eorique, Universit\'e  Paris-Saclay, CEA, CNRS, F-91191 Gif-sur-Yvette Cedex, France}

\begin{abstract}
We analyse  the  dynamics of a light  scalar field responsible for the $\mu$ term of the Higgs potential and coupled to matter  via the Higgs-portal mechanism. We find that this dilaton model is stable under radiative corrections induced by the standard model particle masses.  When the background value of the scalar field is stabilised at the minimum of the scalar potential, the scalar field fluctuations only couple quadratically to the massive fields of the standard model preventing the scalar  direct decay into standard model particles. Cosmologically and prior to the electroweak symmetry breaking, the scalar field rolls down along its effective potential before  eventually oscillating  and  settling down at the electroweak minimum. These oscillations can be at the origin of dark matter due to the  initial misalignment of the scalar field compared to the electroweak minimum, and we find  that, when the mass of the scalar field is less than the eV scale  and acts as a condensate behaving like dark matter on large scales, the scalar particles cannot thermalise with the standard model thermal bath. As  matter couples in a composition-dependent manner to the oscillating scalar, this could lead to a violation of the equivalence principle aboard satellites such as the MICROSCOPE experiment and the next generation of tests of the equivalence principle. Local gravitational tests are evaded thanks to the weakness of the quadratic coupling in the dark matter halo, and we  find that, around other sources,  these dilaton models could be subject to a screening  akin to the symmetron mechanism.

\end{abstract}

\maketitle

\section{Introduction}

Dark matter (DM) is a basic constituent of our standard cosmological model.
A large number of astrophysical observations constrain the amount of such a component up to the percent level. However, there is little  information about its nature. Thanks to measurements of large scale structures, we know that the pressure or kinetic energy of  DM is negligible. In this sense, it is said that DM is cold (CDM). However, there are observations associated with small scales that challenge this standard approach and emphasise potential issues such as the cusp-core problem
\cite{Ostriker:2003qj,Weinberg:2013aya, Pontzen:2014lma,BoylanKolchin:2011de,Moore:1999nt,deBlok:2009sp}.

In order to explore the properties of DM from the point of view of particle physics, it is interesting to study its relation with the electroweak sector. This sector is associated with the mass generation of elementary particles through the Higgs-mechanism. Indeed, the electroweak scale is the highest one in the Standard Model (SM) of particles and interactions, and  also corresponds almost  to the limit of the energy range that can be probed by present particle colliders, i.e. current particle experiments only probe energies slightly larger than the electroweak scale. In addition, this sector suffers from hierarchy problems that need to be understood better from a theoretical point of view, see for instance \cite{Csaki:2015xpj}. In part motivated by these issues, a large number of DM models expressed in terms of weakly interacting massive particles (WIMPs) have been proposed in the last decades. These candidates have the advantage of being produced in the early Universe with the observed order of magnitude for the DM abundance through the so-called freeze-out mechanism. However, no experimental evidence has been found for the existence of WIMPs, see \cite{Feng:2022rxt} for instance for a recent review.

In this work, we consider an alternative DM model directly related to the electroweak sector. In our case, DM arises from the coherent oscillations of a light scalar field responsable for the $\mu$ term of the Higgs potential. In this sense, it has similarities with relaxion models,  discussed in Refs.~\cite{Banerjee:2018xmn,Banerjee:2019epw,Banerjee:2021oeu,Chatrchyan:2022dpy, TitoDAgnolo:2021pjo}. The first proposals about coherent bosonic DM date back to the late seventies in terms of QCD-axion models \cite{Peccei:1977hh,Wilczek:1977pj,Weinberg:1977ma,Vysotsky:1978dc, Preskill:1982cy,Turner:1983sj,Turner:1983he}. In general, a coherent DM framework can be parameterised, from an effective-field-theory point of view, by the DM particle  masses and its self-interactions \cite{Urena-Lopez:2019kud}. The simplest case relies on a massive oscillating scalar field without  self-interactions \cite{Turner:1983he,Sahni:1999qe,Johnson:2008se,Hu:2000ke,Hui:2016ltb}. This type of  generic coherent DM theories are typically indistinguishable from cold DM in relation to  the formation of large-scale structures
\cite{Sakharov:1994id,Sakharov:1996xg,Johnson:2008se, Hwang:2009js, Park:2012ru, Hlozek:2014lca, Cembranos:2015oya,Cembranos:2016ugq}. However, distinctive features at smaller scales can arise from many different causes
 \cite{Hlozek:2014lca,Schive:2014dra, Broadhurst:2018fei,Ostriker:2003qj,Cembranos:2005us,Weinberg:2013aya,Pontzen:2014lma,BoylanKolchin:2011de,Moore:1999nt,deBlok:2009sp,Cembranos:2015oya,Cembranos:2018ulm,Armengaud:2017nkf,Brax:2019fzb,Brax:2019npi,Brax:2020tuk}. Generically, coherent dark matter requires the DM particle mass to be less than the eV scale \cite{Fonseca:2019lmc}. We will see that the eV scale plays a significant role in our scenario, e.g screening of the Sun is only valid for masses larger than $10$ eV.

As the new scalar field determines the Higgs potential, it is coupled to the SM via the Higgs-portal. This fact determines its coupling with the ordinary matter content of the standard model. As such, this light scalar resembles a dilaton field associated with the breaking of conformal symmetry \cite{Goldberger:2007zk}. As we have commented above, our scenario has similarities with relaxion models \cite{Banerjee:2018xmn,Banerjee:2019epw,Banerjee:2021oeu,Chatrchyan:2022dpy} although there are notable differences.  For instance, at the electroweak minimum of the dilaton potential, the linear coupling to matter vanishes and only the quadratic coupling remains. This structure is stable under radiative corrections and the phenomenological signatures change drastically. For instance, we find that the scalar DM cannot thermalise with the standard model bath when the DM mass is lower than the electronvolt scale.

The lack of linear coupling leads to another distinctive feature of the model. Matter couples naturally with a different strength depending on its nature. Therefore, we analyze the possible observational constraints associated with violations of the equivalence principle in experiments such as MICROSCOPE and the prospects for the next generation of experiments. We find that Solar System constraints are evaded due to the weakness of the coupling. But in other environments the quadratic coupling induces  a screening mechanism reducing the constraints on the parameters of the model \cite{Damour:1992kf,Damour:1993id,Cembranos:2009ds,Hinterbichler:2010es}.
In other contexts the scalar field could behave in ways reminiscent of  scalarisation \cite{Damour:1993hw}.
The constraints on  quadratic couplings of ultra light dark matter fields have recently been thoroughly explored in Refs.~\cite{Hees:2018fpg, Banerjee:2022sqg}.

The paper is arranged as follows.
In section~\ref{sec:model}, we introduce the model based on a new scalar degree of freedom. We describe the low energy action, its stability agains radiative corrections and the coupling of the scalar field to the SM. In section~\ref{sec:cosmology}, we analyze the cosmological evolution supported by the dynamics of this scalar field. The main phenomenological consequences of the model are studied in section~\ref{sec:pheno}. In particular, we discuss the violation of the equivalence principle and the E\"otv\"os parameter. Finally, we summarize the main conclusions of our work in section~\ref{sec:conclusion}.

\section{Scalar-dependent $\mu$-term. }
\label{sec:model}

\subsection{The low energy action}

We consider a simple model of electroweak symmetry breaking where one real Higgs field $h$ gives a mass to one Dirac fermion $\psi$. This model is meant to reproduce in a toy-model fashion some aspects of the physics of the electroweak symmetry breaking. We will be interested in the regime where the scalar field $\phi$ is much lighter than the Higgs field $h$, i.e. $m_\phi \ll m_h$.  The Lagrangian of the full theory is given by
\be
{\cal L}= -\frac{1}{2} (\partial \phi)^2 -\frac{1}{2} (\partial h)^2 + \frac{\mu^2(\phi)}{2} h^2 -\frac{\lambda}{4} h^4 -V(\phi)- i \bar \psi \slashed{\partial} \psi - \lambda_\psi h\bar \psi \psi.
\label{lag}
\ee
The electroweak scale is determined by the $\mu^2(\phi)$ mass term  in the broken phase  which depends on the light  field $\phi$.
 As long as $\mu^2(\phi)<0$, the Higgs field does not acquire a vacuum expectation value (vev) and no symmetry breaking occurs. The potential $V(\phi)$ is chosen such that the scalar field induces a change  from values where $\mu^2<0$ to $\mu^2>0$.
 In the following we will assume that as long as $\mu^2<0$ and large, the $\mu^2$ function is mostly linear and the field massless. More precisely, we write the $\mu^2(\phi)$ term as
\be
\mu^2(\phi)= -\Lambda_0^2 + \Lambda^2\frac{\phi}{f} +M^2 \mu_I\left(\frac{\phi}{f}\right) ,
\label{eq:mu}
\ee
where $\mu_I$ is a subdominant term that we can take to vanish at the transition point  $\mu_I(\Lambda_0^2/\Lambda^2)=0$.  The scales $\Lambda_0$,  $\Lambda$ and $M$ are lower than the cut-off scale of the model $\Lambda_c$.
We assume that the following hierarchy is  realised
\be
v\ll  \Lambda_c.
\ee
This corresponds to requiring that the electroweak symmetry breaking happens at low energy compared to the cut-off scale of the theory
{ The scale $f$ determines the dynamics of $\phi$. This could be for instance the vev of a $U(1)$ breaking field if $\phi$ were a pseudo-Goldstone mode.
Notice that we assume that the correction $\mu_I$ is present in the whole range of validity of the effective field theory. In particular, it does not only appear when the electroweak symmetry takes place and we will assume that the effective description is valid from inflation down to lower energies.

 We also assume that the field $\phi$ couples to the inflaton in the Jordan frame through the metric
 \be
 g_{\mu\nu}^J= A^2(\phi) g_{\mu\nu}\,,
 \ee
 where $g_{\mu\nu}$ is the Einstein frame metric. This implies that the scalar potential of the scalar is corrected and becomes \cite{Khoury:2003rn}
 \be
V_{\rm eff}(\phi)= V(\phi) - T [A(\phi)-1]\,,
 \ee
 where we choose
 \be
 A(\phi)= 1 + \frac{(\phi-\phi_e)^2}{2 m^2_{\rm Pl}} \,,
 \label{eq:coupling}
 \ee
and $T$ is the trace of the energy momentum tensor of the inflaton.  During inflation $T=-4 V_{\rm inf}$ where $V_{\rm inf}$ is the potential energy leading to a de Sitter phase. This coupling forces the scalar field to a value $\phi\simeq \phi_e$ at the end of inflation, which then provides an initial condition for the evolution of the field in the post-inflationary universe.  Other mechanisms could be invoked to regulate the early Universe behaviour of the field and slow it down after inflation. Here we consider this simplified description as a proxy for potentially more complex mechanisms which are beyond the scope of this paper, see for instance \cite{Fonseca:2019lmc}. We will discuss the cosmological evolution of the model, including the dynamics of the scalar field during inflation, further in Section \ref{sec:cosmology}.

\subsection{The Higgs phase}

After inflation, the field will evolve until a point where  $\mu^2(\phi)>0$,  and the Higgs field acquires a large mass compared to that of the scalar field $\phi$, as a result one  can `integrate out' the Higgs field using the classical equations of motion
\be
\lambda h^2(\phi)= \mu^2 (\phi) - \frac{\lambda_\psi}{h(\phi)} \bar \psi \psi.
\label{min}
\ee
This method of removing the Higgs degree of freedom is valid as the Higgs-scalar mass matrix does not have a massless eigenvalue \cite{Brax:2021rwk}.
Let us first work at the classical level
 by solving  Eq. (\ref{min}) to lowest order
in a perturbative expansion in $\bar{\psi}\psi$ and obtain
\be
h(\phi)= \frac{\mu (\phi)}{\sqrt{\lambda}} - \frac{\lambda_\psi }{2  \mu^2 (\phi)} \bar \psi \psi \,.
\ee
At lowest order this gives the vev of the Higgs field as
\be
v=\frac{\mu(\bar \phi)}{\sqrt \lambda} \,,
\label{v-vev-def}
\ee
where the dilaton field $\phi$ is stabilised at $\phi=\bar \phi$ with a mass $m_\phi$.
We can also obtain the effective Lagrangian for $\phi$ at the classical level
\be
{\cal L}= -\frac{1}{2} (\partial \phi)^2 -V(\phi) + \frac{\mu^4(\phi)}{4\lambda}  - i \bar \psi \slashed{\partial} \psi - \lambda_\psi \frac{\mu(\phi)}{\sqrt{\lambda}} \bar \psi \psi \,.
\label{eq:efflag}
\ee
This Lagrangian contains the classical part of the potential for $\phi$
\be
V_{\rm clas}(\phi)=V(\phi) -\frac{\mu^4(\phi)}{4\lambda} \,,
\ee
which determines the dynamics of $\phi$ after electroweak symmetry breaking. In the following, we will see that the term in $\mu^4$ can be neglected.

\subsection{Radiative corrections}

A potential for the dilaton field $V(\phi)$ is naturally present as it can be   induced by radiative corrections of the Higgs field $h$. Closing the Higgs loop in the coupling provided by Eq. \eqref{lag}, i.e. ${\mu^2(\phi)}h^2/2$, gives :
\be
V_{\rm one \, loop}(\phi)\supset \frac{\Lambda_c^2}{32\pi^2}{\mu^2(\phi)}\,,
\label{V1loop}
\ee
where $\Lambda_c$ is the scale at which the Higgs quadratic divergence gets cut off. In the spirit of effective field theories where allowed couplings should be present, we will assume that the potential for the scalar $\phi$ 
is corrected at the one loop level by the Higgs loop to
\be
V(\phi)= V_0(\phi) -  a \frac{\Lambda_c^2}{32\pi^2}{\mu^2(\phi)} \,,
\ee
where $a$ is a dimensionless constant. We will assume that $a>0$ in order to ensure that this contribution to the scalar potential decreases from the symmetric phase to the electro-weak breaking one. We will also simplify the model by taking $V_0(\phi)=V_0$ which makes sure that the vacuum energy vanishes at the minimum of the potential.

There are also logarithmic corrections to the potential.
As the mass of the Higgs field  is
$m^2_h= 2 \mu^2(\phi)$ when the electroweak symmetry is broken, the
corrections to the scalar potential  are proportional to $\mu^4(\phi)$. At the one loop level this yields for the total effective potential
\be
V_{\rm eff}(\phi)=V(\phi) -\frac{\mu^4(\phi)}{4\lambda}+ \frac{\mu^4(\phi)}{16\pi^2} \ln \frac{\Lambda_c^2}{2\mu^2(\phi)} \,,
\label{corrH}
\ee
where $\Lambda_c $  is the UV renormalisation scale of the Higgs-scalar theory. This is the usual Coleman-Weinberg correction at one loop calculated in dimensional regularisation \cite{Coleman:1973jx}. The corrections due to the masses of matter particles have the same form. Indeed, the masses are all proportional to $\mu (\phi)$ and therefore lead to the same type of one-loop corrections. Higher order loops should also contribute to the effective potential in $\mu^4$ as $\mu$ is the only mass scale in the theory below the cut-off. In Eq.~(\ref{corrH}) we see that the corrections in $\mu^4$ can be incorporated in a redefinition of the
self-coupling $\lambda$ which becomes $\lambda (v)$ due to the logarithmic corrections
\be
\lambda \to \lambda(v)= \lambda \left(1+ \frac{\lambda}{4\pi^2}  \ln \frac{\Lambda_c^2}{2\lambda v^2}\right)\,,
\ee
and similarly for loops coming from matter fields.
Loops induced by the scalar field itself will scale in $m_\phi^4$ which is assumed to be very small compared to $\mu^4$. Hence we neglect the self-loops of $\phi$ in the following and work at the classical level when it comes to the scalar field $\phi$. The self-coupling will always be taken to be the renormalised one $\lambda (v)$ at the electro-weak scale.

In the following, we will focus on a scenario where the scalar field $\phi$ evolves cosmologically towards the minimum $\bar\phi$
of the scalar potential and eventually oscillates around this extremum, such that oscillations are described by $\varphi=\phi-\bar\phi$. In particular we will focus on models where
\be
V(\phi)=-g\Lambda_c^2{\mu^2(\phi)}+ V_0\,,
\label{eq:effpot}
\ee
where $g=
 \frac{a}{32\pi^2}>0$. The minimum of the scalar potential in the radiation era
 corresponds
to
\be
\partial_\phi \mu^2 (\bar \phi)=0 \,.
\ee
The contribution in $-\mu^4/4\lambda$ from the Higgs phenomenon does not change this result and is always negligible as $\mu \ll \Lambda_c$
close to the electro-weak transition.

\subsection{Couplings to bosons}
So far we have only considered the coupling of the scalar $\varphi$ to fermions via the Higgs portal. Couplings to photons and gluons are induced by triangular anomalous diagrams where massive fermions run in the loop. This affects the low energy theory of the dark matter field each time a  particle of the standard model decouples \cite{Goldberger:2007zk}.  This  induces an effective interaction Lagrangian of the type
\be
{\delta L}= - \frac{\alpha_F(E) e^2}{4} \frac{\varphi^2}{\Lambda^2_f} F_{\mu\nu}F^{\mu\nu} - \frac{\alpha_G(E)g^2_3}{4} \frac{\varphi^2}{\Lambda^2_f} G_{\mu\nu}G^{\mu\nu}\,,
\label{eq:bosons}
\ee
where $\alpha_{F,G}(E)$ are numerical constants depending on the charges of the decoupled fermions, i.e. fermions more massive than $E$,  under the electromagnetic $U(1)$ symmetry and the QCD (Quantum ChromoDynamics) $SU(3)$ group, see for instance \cite{Goldberger:2007zk,Brax:2010uq} for an explicit discussion\footnote{The coefficients are given by $\alpha_F(E)= \frac{\sum_{i=1}^{N_f^>} q_i^2}{24\pi^2}$, where $N_f^>$ is the number of particles and antiparticles of charges $q_i$ which have decoupled at the energy $E$. Similarly we have
$\alpha_G(E)= \frac{\sum_{i=1}^{N_f^>} T(R_i)}{12\pi^2}$, where the fermions are in the representation $R_i$ such that the Lie algebra generators are normalised by ${\rm Tr}(T^a T^b)= T(R_i) \delta^{ab}$.}. The coupling constant $e$ and $g_3$ are the electromagnetic and QCD coupling respectively evaluated at the energy scale $E$. The sign of the interaction Lagrangian is given by the sign of the interaction between the scalar and fermions.
In the following we will be interested in the low energy effects of the coupling to photons at energies well below the electron mass. As a result, the coefficient $\alpha_F$ will take into account the decoupling of all the standard model particles. On the other hand, the coupling to the gluons is relevant to determining the QCD condensation scale. In this case, $\alpha_G$ depends on the decoupling of the heavy quarks $c,b$ and $t$. We will return to the consequences of these couplings in Section \ref{sec:conse}.

\subsection{Back-reaction}

Using the effective Lagrangian, Eq.~(\ref{eq:efflag}), we can see that in the Higgs phase, when standard model fermions have acquired a mass, the scalar potential is modified by the average fermion number
\be
V_{\rm matter}(\phi)=V_{\rm eff}(\phi)+ \frac{\lambda_\psi}{\sqrt \lambda}\mu (\phi) n_\psi \,,
\ee
where $n_\psi= \langle \bar \psi \psi\rangle$ is the fermion number density. This is only valid when $\mu^2 >0$, and we are in the Higgs phase. When $\mu^2<0$, the matter effect disappears and the potential is simply
\be
V_{\rm no \ matter}(\phi)=V_{\rm eff}(\phi) \,.
\ee
This back-reaction behaviour is reminiscent of a coupling to the trace of the energy momentum tensor for scalar-tensor theories \cite{Khoury:2003rn}.
Close to the electroweak transition, the matter term is much smaller than the effective potential as $n_\psi\sim T^3$ where $T\sim v$ and $\Lambda_c\gg v$. This implies that in the vicinity of the electroweak transition, the matter corrections are negligible.

In addition to the coupling in Eq.~(\ref{eq:efflag}), we could  included higher order couplings between the Higgs field and matter of the type
\be
\delta {\cal L}\subset - \frac{h^n}{\Lambda_h^{n-1}}\bar\psi \psi \,,
\label{corrr}
\ee
where $\Lambda_h$ is a cut-off scale in the Higgs sector. Typically this type of operator leads to a matter correction to the scalar potential
\be
\delta V_{\rm matter}\subset \frac{v^n}{\Lambda_h^{n-1}} n_\psi \,,
\ee
after the electroweak transition. This is always a negligible contribution to the potential at the electroweak scale for $T\sim v$ as $v\ll \Lambda _h\lesssim \Lambda_c$.

In the following, we will consider models where the field $\phi$ can escape the vicinity of the minimum at $\bar\phi$  if the field $\phi$ reaches values such as $\mu^2 \simeq -M^2$ where $M$ is a large scale taken to be smaller than the cut-off scale $\Lambda_c$.  Now the matter back-reaction would stop the field from jumping over the barrier associated to the potential  $V(\phi)=-g\Lambda_c^2 \mu^2$ if the $\mu$ term were prevented from reaching a large value of order $M$. We assume that the scale $M$ is the natural scale $M\gg v$ of $\vert \mu (\phi)\vert $ far away in field space from the minimum of the potential where $\mu(\bar \phi) \ll M$. In this case, this stopping mechanism would be  reminiscent of the Damour-Nordtvelt effect \cite{Damour:1992kf,Damour:1993id} for the models considered here, whereby the electroweak scale should be driven close to the zero of $\mu$ cosmologically in an attractor fashion.
The higher order terms in Eq.~(\ref{corrr}) provide a stopping correction to the scalar potential  of the form
\be
\delta V(\phi) \simeq \frac{\mu^n(\phi)}{\Lambda_h^{n-1}}v^3 \,,
\ee
where $h(\phi)=\mu(\phi)/\sqrt \lambda$ and $n_\psi \simeq v^3$.
This term dominates compared to $V(\phi)$  when $\mu \sim M$ provided
\be
\left(\frac{M}{\Lambda_h}\right)^{n-2}\gtrsim \frac{\Lambda_c^2 \Lambda_h}{v^3}\,.
\ee
This can be
realised if $M\gg \Lambda_h$ where higher and higher corrections to the Higgs portal would back-react strongly on the dynamics of the dilaton. Of course this is beyond the realm of the effective field theory set up we have adopted here as the full non-perturbative series has to be known. The only conclusion we can draw is that a full analysis of the Higgs sector and its coupling to matter is necessary to probe the large $\mu^2>0$ region of the theory. In particular, it is quite likely  that after the electroweak phase transition the effects of such matter couplings  could be  efficient enough to stop the dilaton and guarantee that the field simply oscillates around the minimum where $\mu\sim v$ close to $\mu=0$. A full discussion of this issue goes beyond the present paper, and we refer the reader to Ref.~\cite{Fonseca:2019lmc}. In addition to the higher order corrections to the Higgs coupling, thermal effects must be taken into account, as it has been discussed in Ref. \cite{Cembranos:2009ds}. As the coupling of the scalar field is proportional to the trace of the energy-momentum tensor, it is commonly assumed to vanish during the radiation dominated epoch in the early universe. However, finite radiative corrections to the  coupling impact generically on the evolution of the field, modifying the allowed region of its parameter space, changing the abundance of different cosmological relics and producing early phases of contracting evolution (in the Jordan frame) \cite{Cembranos:2009ds}. In the following, we will analyse the dynamics of the theory by neglecting these thermal effects, that will be taken into account in future works.

\subsection{The coupling to matter}

The light scalar field $\phi$ couples to matter via the Higgs portal.  In the Higgs phase where matter fields acquire a mass, one can
expand the light scalar field  $\phi= \bar \phi +\varphi$ to obtain the  resulting interaction between $\varphi$ and matter
\be
{\cal L}_{\rm int}= -\frac{\beta}{m_{\rm Pl}} m_\psi \varphi\bar \psi \psi - \frac{m_\psi}{2\Lambda_f^2} \varphi^2\bar \psi \psi \,,
\label{quadr}
\ee
 where we have identified $m_{\psi}= \lambda_\psi v$. The scalar has a  Yukawa type coupling
\be
\beta\equiv m_{\rm Pl}  \frac{\partial_\phi \mu(\phi)\vert_{\bar \phi}}{\sqrt{\lambda}v}=0\,
\label{beta}
\ee
where the Higgs vev $v$ was defined in Eq. (\ref{v-vev-def})
and a quadratic coupling
\be
\frac{1}{2\Lambda_f^2}=  \frac{\partial_\phi^2 \mu}{2\mu} = - \frac{m^2_\phi }{4 g \mu^2 \Lambda_c^2}  \,,
\ee
which is composition independent at the fundamental fermion level, i.e. it depends not on the fermion species $\psi$. On the other hand, the coupling to nucleons will depend on the species and this could induce violations of the equivalence principle. We will investigate this possibility below.

As a result the light scalar field $\varphi$ whose mass is given by
\be
m^2_\phi = - \frac{\mu^2\partial_\phi^2\mu^2}{2\lambda} + V''\vert_{\bar\phi}\,,
\label{massphi}
\ee
is stable quantum-mechanically, i.e. there is no decay into two fermions at tree level in vacuum. The scalar mass is dominated by
\be
m^2_\phi \simeq -g \Lambda_c^2 \partial^2_\phi \mu^2 =-2 g \Lambda_c^2 \mu \partial_\phi^2\mu\,,
\ee
where the last equality is valid at the minimum where $\partial_\phi \mu=0$. We will see below how this mass scale can be much smaller than $m_h$.

In the following we will be interested in $\varphi$ as a candidate for dark matter where $\varphi$ is light, i.e. $m_\phi \lesssim 1$ eV \cite{Fonseca:2019lmc}. In principle the coupling Eq.~(\ref{quadr}) could induce  the thermal equilibrium between $\phi$ and the standard model fields, eventually leading to the freezing out of the $\phi$ abundance which could then be adjusted to match the present amount of dark matter in the Universe. But as we will see in Section \ref{sec:therm}, it is not possible for the scalar field in our model to be in thermal equilibrium with the thermal bath.
Another possibility, as $m_{\phi}$ is small,  could be  that the scalar $\phi$ decoupled when relativistic. The remaining abundance of hot dark matter behaves as $\Omega_\phi h^2 \simeq 10^{-3} (\frac{100}{g_\star}) (\frac{m_\phi}{{\rm 1 \ eV}})$, where $g_\star$ is the number of relativistic species at decoupling. For very light scalars, the abundance of hot dark matter becomes negligible.

So the scalar field in our model can only play the role of dark matter in a non-thermal fashion and decouples from quarks and leptons at the electroweak symmetry breaking. In this case, decoupling happens before the scalar acquires a mass and therefore causes no issue with the abundance of dark matter, whether hot or cold.
In the following we will consider the natural situation where the scalar field rolls down along its potential from small values before oscillating around its minimum. This mechanism is similar to the misalignment mechanisms  used for axion \cite{Abbott:1982af,Dine:1982ah,Preskill:1982cy,Turner:1983he,Marsh:2015xka} or scalar dark matter models such as fuzzy dark matter \cite{Hui:2016ltb}. The abundance of dark matter in these cases is related to the amplitude of the oscillations around the minimum.

\subsection{An explicit dilaton model}

As we expect fluctuations of the scalar around the minimum of its potential to play the role of dark matter, we now return to finding the minimum of the scalar potential and imposing that electroweak symmetry breaking takes place at the scale $v$. For this, let us notice that in the radiation dominated era
the minimum of the effective potential, Eq.~(\ref{eq:effpot}) with $\mu$ defined in Eq.~(\ref{eq:mu}),   is such that
\be
\left.\frac{d \mu_I (y)}{dy}\right\vert_{\bar y}= -\frac{2\Lambda^2}{\pi M^2} \,,
\label{mimi}
\ee
in terms of the rescaled scalar field $y=\pi \phi/(2f)$ and  its  vev  $\bar y$.
The electroweak scale is determined by imposing that at the minimum we have $\mu^2 (\bar\phi) = \lambda v^2$ which implies that
\be
\mu_I(\bar y)-\bar y \left.\frac{d\mu_I}{dy}\right\vert_{\bar y}= \frac{\Lambda_0^2}{M^2}+ \lambda \frac{v^2}{M^2}\,.
\label{tune}
\ee
As a typical example
{we choose a correction to the $\mu^2$ term of the axion type where
\be
\mu_I (y)= \cos y\,.
\ee
where we have assumed that $\Lambda_0=\Lambda$.
The axionic contribution vanishes for $y=\pi/2$ so that the transition from $\mu^2<0$ to $\mu^2>0$ takes place where both the linear part of $\mu^2$ and $\mu_I$ vanish.
Explicitly we have
\be
\mu^2(\phi)= \Lambda^2 \left(\frac{\phi}{f}-1\right) +M^2\cos \left( \frac{\pi}{2}\frac{\phi}{f} \right)\,.
\label{eq:musquared}
\ee
We find that Eq.~(\ref{mimi}) implies that
\be
\sin  \bar y= \frac{2}{\pi} \frac{\Lambda^2}{M^2}\,.
\label{eq:symm}
\ee
The tuning of the electroweak symmetry breaking, Eq.~(\ref{eq:symm}), can be satisfied provided
\be
\cos\bar y + \bar y \sin \bar y=  \lambda \frac{v^2}{M^2} + \frac{\Lambda_0^2}{M^2}\,.
\ee
This can be easily analysed by expanding the $\mu^2$ term around $y=\pi/2$ using $\bar y=\pi/2(1  + \delta)$. This gives
\be
\mu^2(\phi)= \left(\Lambda^2 -\frac{\pi}{2}M^2\right)\delta + \frac{M^2}{6}\left(\frac{\pi}{2} \delta\right)^3.
\ee
The minimum of the potential is then
given by
\be
\delta^2= 2\left(\frac{2}{\pi}\right)^3 \left(\frac{\pi}{2}-\frac{\Lambda^2}{M^2}\right)\,.
\label{eq:delta1}
\ee
Imposing that $\mu^2(\bar \phi)=\lambda v^2$ implies that
\be
\delta= \frac{3}{2}\frac{\lambda v^2}{\Lambda^2 -\frac{\pi}{2}M^2}\,.
\label{eq:delta2}
\ee
These two conditions, Eqs. (\ref{eq:delta1}) and (\ref{eq:delta2}), are compatible provided the two scales $\Lambda$ and $M$ are related by
\be
\frac{\Lambda^2}{M^2}=\frac{\pi}{2} \left[1- \left(\frac{3\lambda v^2}{2\sqrt 2 M^2}\right)^{2/3}\right]\,.
\label{eq:compatible}
\ee
This may appear as a tuning of the potential although we have seen that radiative corrections preserve the shape of $\mu^2$ and therefore the ratio between $\Lambda$ and $M$.

Defining
\be
\epsilon= \left(\frac{3\lambda v^2}{2\sqrt 2 M^2}\right)^{2/3}\,,
\label{eps-def}
\ee
we find that the value of $\delta$ at the minimum of the potential is
\be
\frac{\pi}{2}\delta_{\rm min}= -\sqrt 2 \left(\frac{3\lambda v^2}{2\sqrt 2 M^2}\right)^{1/3}= -\sqrt{2\epsilon}\,.
\ee
Notice that this extremum is a maximum of $\mu^2$. As a result we get a minimum for the scalar potential and a mass for the scalar field
\be
m^2_\phi= -g \left(\frac{\pi}{2}\right)^3 \delta_{\rm min} \frac{M^2 \Lambda_c^2}{f^2}\,.
\ee
This mass is reduced compared to the naive expectation of the mass that one might obtain from considering Eq.~(\ref{eq:musquared}) and assuming $\cos(\pi \phi/2f) \sim 1$, which would give
\be
m_0^2= g\frac{\pi^2}{4}\frac{M^2 \Lambda_c^2}{f^2}.
\ee
In fact the mass is reduced by  a factor of $\delta_{\rm min}$ \cite{Banerjee:2018xmn}, i.e.
\be
m^2_\phi= -\frac{\pi}{2} \delta_{\rm min}m_0^2= \sqrt{2\epsilon}~ m_0^2\,.
\ee
{ The mass $m_0$ is suppressed compared to the cut-off scale by a factor of $f$. Moreover the mass of the scalar field $m_\phi$ in the shallow part of the potential close to the minimum  is reduced a factor of $\epsilon^{1/4}$ which is also small, see Fig. \ref{fig:plot-V-y}. This small mass will eventually be identified with the mass of the scalar dark matter in the Universe.}
We are also interested in the location of the field where the potential vanishes. The potential vanishes where $\mu^2=0$ corresponding to  $\delta=0$ and also for

\be
\frac{\pi}{2}\delta_0= -\sqrt 6 \left(\frac{3\lambda v^2}{2\sqrt 2 M^2}\right)^{1/3}=-\sqrt{6\epsilon}\,.
\ee
In the interval between $\delta_0$ and the origin, the mass of the field is of order of  $m_\phi$. In this region, the potential is very flat as can be seen in Fig. \ref{fig:plot-V-y}.

\section{Cosmological Evolution}
\label{sec:cosmology}
\subsection{Inflation}

During inflation, the electroweak symmetry is preserved and $\langle h\rangle =0$. The dynamics of $\phi$ are determined by
\be
V_{\rm eff}(\phi)= -g\Lambda_c^2 \mu^2 (\phi) + 6H^2_{\rm inf}{(\phi-\phi_e)^2} +V_0\,.
\ee
where the form of the coupling to the inflaton Hubble rate, $H_{\rm inf}$, is assumed to be given in Eq.~(\ref{eq:coupling}).
The large quadratic term due to the coupling to inflation  forces $\phi$ to be close to $\phi_e$ and stabilised.
The minimum of the effective potential is obtained for
\be
\phi_{\rm inf}= \phi_e+ \frac{g\Lambda_c^2}{12 H^2_{\rm inf}}\partial_\phi \mu^2\vert_{\phi=\phi_{\rm inf}}\,.
\ee
This is of order
\be
\frac{\phi_{\rm inf}-\phi_e}{f}\propto \frac{{g}\Lambda_c^2 M^2}{f^2 H^2_{\rm inf}}\sim \frac{m^2_0}{H^2_{\rm inf}}\ll 1\,.
\ee
which is very small as long as we assume that {$ m_0 \ll H_{\rm inf}$, and in addition we will assume that $H_{\rm inf}\ll \Lambda_c$.}
In particular this implies that the field does not move at the end of inflation until the Hubble rate goes down to $H\simeq m_0$ { in the post-inflationary era}, an expectation that is confirmed by our numerical solutions, see Figure \ref{cosmo1}. {This is independent of the choice of  $\phi_e$.} { Notice that during inflation the mass of the scalar field is $m^2_{\rm inf}= 12 H^2_{\rm inf}$ implying that the field is heavy and no isocurvature fluctuations are generated.}

\subsection{Post-inflation evolution}

\begin{figure}[t]
\includegraphics[width=0.7\textwidth]{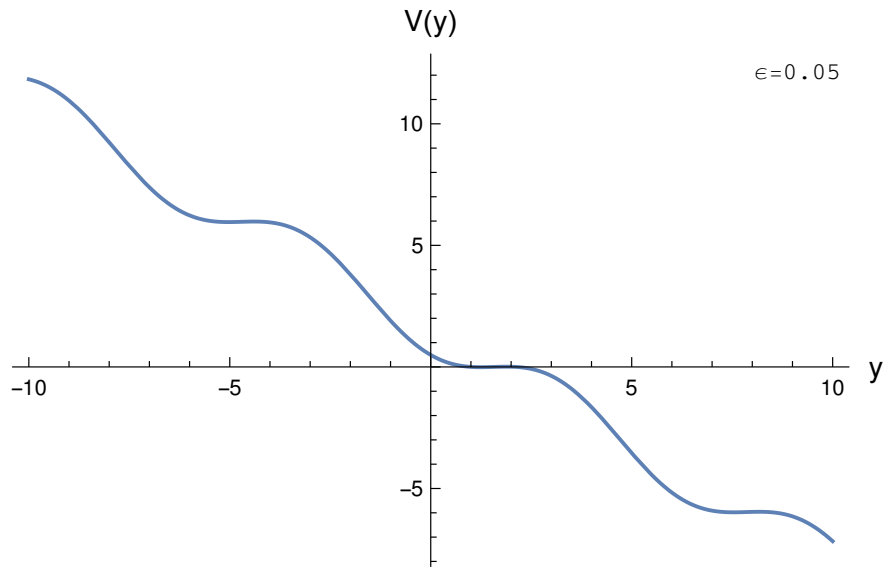}
\centering
\caption{\small The normalised  potential $V(y)$ where $y=\frac{\pi\phi}{2f}$ for $\epsilon=0.05$. The oscillations of the dark matter field take place on the flat part of the potential close to the first minimum on the positive real axis. The field is first stabilised during inflation and then released in the post-inflationary era when the Hubble rate drops below the mass of the scalar field. The oscillatory behaviour is guaranteed as long as the potential is not too flat. The mass on the steep part of the potential is typically $1/\sqrt \epsilon$ larger than close to the minimum.  }
\label{fig:plot-V-y}
\end{figure}

\begin{figure}[t]
\includegraphics[width=0.8\textwidth]{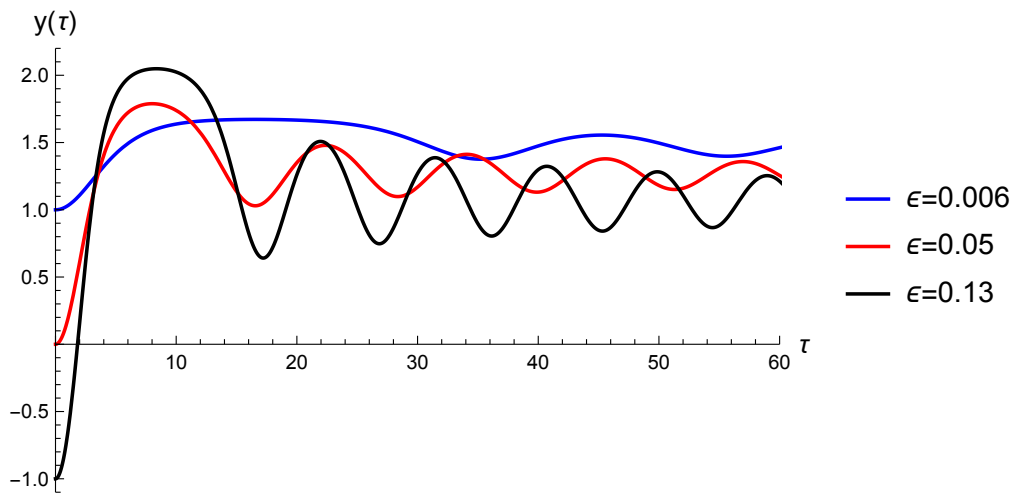}
\centering
\caption{\small Scalar field trajectories $y(\tau)=\frac{\pi\phi}{2f} $ as a function of $\tau$ for the initial conditions $y_0 = 1$, $0$, $-1$, and the smallest associated values of $\epsilon$ that ensure the field remains trapped in the first local minimum.}
\label{fig:y-tau}
\end{figure}

The equation of motion of the scalar field is the Klein-Gordon equation
\be
\ddot\phi + 3 H \dot\phi + \frac{dV}{d\phi} = 0\,,
\ee
with the specific potential
\be
V(\phi) = - g \Lambda_c^2 \mu^2 +V_0= - g \Lambda_c^2 \left[ \Lambda^2 \left( \frac{\phi}{f} - 1 \right) + M^2 \cos\left(\frac{\pi}{2} \frac{\phi}{f} \right) \right] +V_0.
\label{eq:V-phi}
\ee
It is useful to change the time variable to
$ \tau = m_0 t ,
$
to obtain the reduced equation in the radiation era
\be
y'' + \frac{3}{2\tau} y' + \sin(y) = 1 - \epsilon \,,
\ee
where, as before, $y=\pi \phi/(2f)$,  $\epsilon$ was defined in Eq.(\ref{eps-def}) and the primes denote derivatives with respect
to the rescaled time $\tau$.
This corresponds to the motion of the particle $y(\tau)$ in the potential
$V(y)=-(1-\epsilon) y - \cos(y)$ shown in Fig.~\ref{fig:plot-V-y}.

The field starts rolling at the end of inflation, at time $\tau_i \ll 1$, with zero velocity.
In the limit $\tau_i \to 0$, there is a regular solution with a Taylor expansion of the form
\be
y(\tau) = y_0 + y_2 \tau^2 + y_4 \tau^4 + y_6 \tau^6 + \dots
\ee
Substituting into the equation of motion gives the coefficients
\be
y_2 = \frac{1-\epsilon-\sin(y_0)}{5} , \;\;\; y_4 = \frac{\cos(y_0) [-1+\epsilon+\sin(y_0)]}{90} , \;\; \dots
\ee
and we can see that the field only moves significantly after a time $\tau \gtrsim 1$, that is,
after $H \lesssim m_0$ \footnote{A more detailed study of the slow roll evolution of the field after the end of inflation can be found in appendix \ref{app:rad}.}.

In this paper, we assume that the field starts near the first positive minimum, located
at $y \simeq \pi/2$, that is, $-1 \lesssim y_0 \lesssim \pi/2$.
Then, for $\epsilon > 0$ not too small, because of the Hubble friction, the field will remain
trapped inside this first shallow potential well and oscillate at late times around the equilibrium
value
\be
\bar y_{} = \arcsin(1-\epsilon) = \frac{\pi}{2} - \sqrt{2\epsilon} + \dots , \;\;\;
V''(\bar y_{}) = \sqrt{2\epsilon} + \dots
\ee
We show in Fig.~\ref{fig:y-tau} the trajectories obtained for different values of the initial
condition $y_0$ with, in each case, the smallest value of $\epsilon$ that keeps the field
trapped in the first local minimum.
We see that as $y_0$ is decreased from $\pi/2$,
the parameter $\epsilon$ must increase to enhance the barrier at the right of the local
minimum.
However, thanks to the Hubble friction, which slows down the rolling down
the potential, small values of $\epsilon \lesssim 0.1$ are sufficient to keep the field in the
local potential well for a reasonably large range of initial conditions, $-1 \lesssim y_0 \lesssim 1$. From the definition of $\epsilon$ in equation (\ref{eps-def}) we see that this is guaranteed as long $M\gtrsim 1$ TeV. For larger values of $M$, $\epsilon$ would be smaller and as a result the basin of attraction of the minimum would shrink.

Writing $y=\bar y_{}+\frac{\pi}{2}\delta$, as before, we obtain for the late-time small oscillations the equation of motion
\be
\delta'' + \frac{3}{2\tau} \delta' + \sqrt{2\epsilon} \, \delta = 0 \,.
\ee
This gives decaying oscillations of the form $\delta =\tau^{-1/4} J_{\pm 1/4}( (2\epsilon)^{1/4} \tau)$.
As $\tau \gg 1$ this corresponds to harmonic oscillations with an amplitude that slowly decays
as $\tau^{-3/4} \propto a^{-3/2}$, where $a(\tau)$ is the cosmological scale factor.
This gives an energy density $\rho_\phi$ that decays as $a^{-3}$, as for dark matter.
The potential $V(\phi)$ of Eq.(\ref{eq:V-phi}) reads
$V(\bar \phi_{}) + g \Lambda_c^2 M^2 \sqrt{\epsilon/2} (y-\bar y_{})^2+\dots$
The first local maximum for $y> \bar{y}$  is for $y_{\max} = \pi/2+\sqrt{2\epsilon} + \dots$.
If the scalar field first turns around at a value smaller but of the order of $y_{\max}$,
we obtain that at $\tau \sim 1/2$ the dark matter density is
\be
H \sim m_0 : \;\;\; \rho \sim g \Lambda_c^2 M^2 \epsilon^{3/2} \sim g \lambda \Lambda_c^2 v^2 \,.
\ee
At later times this energy density evolves like cold dark matter and decays as $a^{-3}$.  The initial dark matter density  is then
\begin{equation}
    \rho_{\rm in} = \frac{1}{2} m_{\phi}^2 \phi_0^2\,,
\end{equation}
which is then red-shifted up to now to become the dark matter density in the present Universe $\rho_0$.

 \subsection{Thermalisation}
\label{sec:therm}

We can now come back to the thermalisation of the scalar particles and impose that  the scalar is never in equilibrium with the particles of the Standard Model. This requires that the decoupling temperature
$T_{\rm dec}\gtrsim v$. Let us examine this scenario.

The coupling between the scalar and matter is quadratic in the scalar with a coupling constant $1/\Lambda_f^2$ of order  $ {m^2_\phi} /{v^2\Lambda_c^2}$. In the relativistic regime, the square of the scattering amplitude $ \phi+\phi \to \psi +\psi$ behaves like \footnote{{Note that in this expression $v$ is a spinor, whereas elsewhere in this article $v$ refers to the vacuum expectation value of the Higgs field.}}
\be
\vert {\cal M}\vert^2 \simeq \frac{m_\psi^2}{4\Lambda_f^4} (u\bar v)^2\,,
\ee
where in the relativistic regime the external spinors are such that $u \bar v\simeq T$.
The cross section is of order
\be
\sigma \simeq \frac{ \vert {\cal M}\vert^2}{s} \,,
\ee
where the Mandelstam variable $s\simeq T^2$, which yields
\be
\sigma \simeq \frac{m_\psi^2}{128\pi^2\Lambda_f^4} \,,
\ee
which is constant\footnote{The cross section is given by $\sigma= \frac{ m^2_\psi}{128\pi^2\Lambda_f^4} \frac{s-4m_\psi^2}{s} \sqrt{\frac{s-4m^2_\psi}{s-4m^2_\phi}}$. Taking the limit $s\sim T \gtrsim m_\psi, m_{\phi}$, we find $\sigma= \frac{ m_\psi^2}{128 \pi^2\Lambda_f^4}$.}.
In this regime the number of relativistic species is given by $\tilde g_\star$ and the
the number density of massive particles reads $n\simeq \tilde g_\star \zeta(3) T^3/\pi^2$ where $\tilde g_\star= \sum_{\rm bosons}g_i^B + \sum_{\rm fermions} \frac{3}{4} g_i^F $ and $g_i^{B,F}$ are the degeneracy factors for bosons and fermions. The reaction rate is therefore
\be
\Gamma= \frac{\tilde g_\star\zeta(3)}{\pi^2} \sigma T^3\simeq \frac{m_\psi^2}{4\Lambda_f^4}  T^3 \,,
\ee
and the Hubble rate is given by
\be
H\simeq \sqrt {\frac{g_\star\pi^2}{30}}  \frac{T^2}{m_{\rm Pl}}\,,
\ee
where $g_\star= \sum_{\rm bosons}g_i^B + \sum_{\rm fermions} \frac{7}{8} g_i^F$. The sum is taken over all the particles of the Standard Model.
Equilibrium is maintained when $\Gamma \gtrsim H$ which implies
\be
T\ge T_{\rm dec}\simeq \frac{4\Lambda_f^4}{ m_\psi^2 m_{\rm Pl}} \,,
\ee
where we have taken $\tilde g_\star \sim g_\star \sim 100$.

The scalar is never in thermal equilibrium when $T_{\rm dec}\gtrsim v$. { {We use this criterion as at higher temperatures the electroweak transition has not occurred and the {dilaton} field does not behave like dark matter.}}
This corresponds to
\be
\frac{m_\psi}{2\Lambda_f^2} \lesssim (m_{\rm Pl}v)^{-1/2}\,,
\label{bm}
\ee
or more appropriately
\be
\frac{m_\phi}{v} \lesssim \frac{\Lambda_c}{m_{\rm Pl}}\left(\frac{m^3_{\rm Pl}}{m_\psi^2 v}\right)^{1/4}\simeq 10^{13} \frac{\Lambda_c}{m_{\rm Pl}}.
\label{eq:mphiv}
\ee
{where the final approximate equality arises when $m_{\psi}\sim \mbox{ GeV}$ corresponding to the $b$ quarks.} Indeed, the most severe constraint comes from the heaviest quark with a mass $m_\psi$ of a few GeV's. As the cut-off scale, $\Lambda_c$,  must be larger than one  TeV, the bound in Eq.~(\ref{eq:mphiv}) is always satisfied as soon as  $m_\phi \lesssim 1$ GeV. This is always satisfied, as  the range of masses for which the occupation number of the oscillating scalar is large enough to describe dark matter is $m_\phi\lesssim 1$ eV \cite{Brax:2020oye}. As a result we conclude that the oscillating dilaton scalar never thermalises with the standard model and can describe dark matter via the misalignment mechanism.

\begin{figure}[t]
\includegraphics[width=0.6\textwidth]{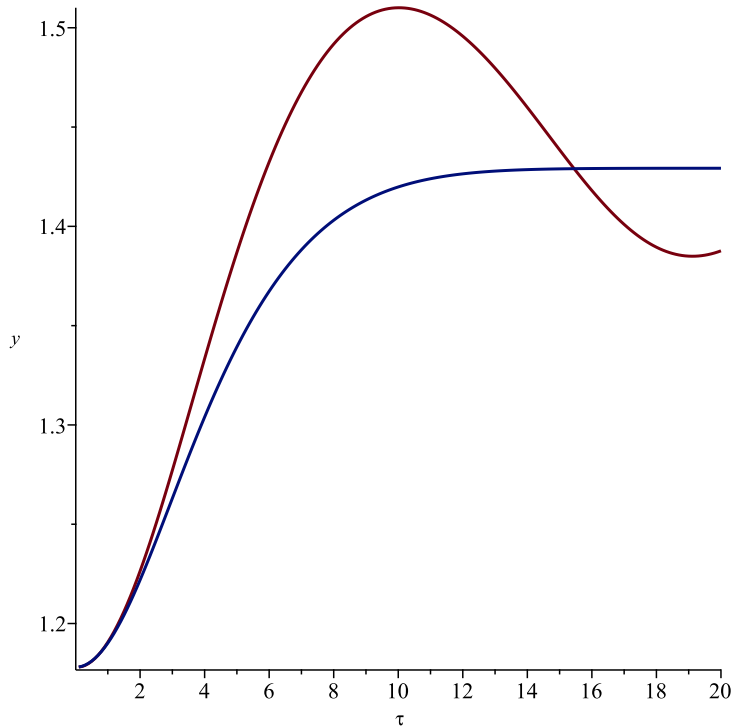}
\centering
\caption{\small The normalised $y=\pi\phi/2f$  field as a function of $\tau$ for $\epsilon=0.01$. The blue curve is the numerical solution. The red one the approximate solution. One can see that the slow roll approximation is valid for a few Hubble times. }
\label{cosmo1}
\end{figure}

\section{Violation of the equivalence principle}
\label{sec:pheno}
\subsection{The coupling to matter}
The oscillating scalar field around the minimum of the $\varphi$ scalar potential induces a quadratic coupling to fermions of the type
\be
{\cal L}\supset \frac{m_\psi}{2\Lambda_f^2} \varphi^2 \bar\psi \psi \,,
\ee

which implies a universal dependence of the fermion masses
\be
m_\psi (\varphi)= m_\psi \left(1- \frac{\varphi^2}{2\Lambda_f^2}\right).
\ee
Notice that the coupling to matter tends to destabilise the scalar field like in scalarisation models \cite{Damour:1993hw}.
As such the  weak equivalence principle is respected at the level of elementary particles as each fermion couples universally to the dilaton with a Jordan frame metric
\be
g_{\mu\nu}^\psi= \left(1- \frac{\varphi^2}{2\Lambda_f^2}\right)^2 g_{\mu\nu}\,,
\ee
where $g_{\mu\nu}$ is the Einstein frame metric.
On the other hand, as macroscopic matter is composed of atoms themselves comprising a nucleus and electrons, the coupling to a particular species  depends on the number of nucleons $A$ and the number of electrons $Z$ of the material. This is important as tests of the equivalence principle are carried out with two different bodies with different numbers of electrons and nucleons.
The Klein-Gordon equation for the scalar $\varphi$ in the presence of matter species $A$ of density $\rho_A$ now reads
\be
\Box \varphi= m^2_\phi \varphi + \frac{\alpha_A (\varphi)}{m_{\rm Pl}} \rho_A\, .
\ee
The species-dependent coupling function is defined as
\be
\alpha_A(\varphi)= m_{\rm Pl}Q_A \frac{\partial_\varphi m_\psi(\varphi)}{m_{\psi}} = -Q_A \frac{m_{\rm Pl}}{\Lambda_f^2}\varphi \,,
\label{couling}
\ee
where $Q_A$ is a dimensionless phenomenological coefficient. Concentrating on the contributions from the fermion masses, $Q_A$ is given by
\be
Q_A= Q+ [Q'_A]_{\hat m}+ [Q'_A]_{ m_e}+[Q'_A]_{\delta m}\,,
\label{QQ}
\ee
where the couplings to the average mass of the $u$ and $d$ quarks is $[Q'_A]_{\hat m}$, the coupling to the electron mass is $[Q'_A]_{ m_e}$ and the coupling to the $u$-$d$ mass difference is $[Q'_A]_{\delta m}$. They are tabulated for different metals \cite{Hees:2018fpg}. The universal coupling $Q\simeq 0.093$ is also phenomenological. Depending on the sign of $Q_A$, the model will behave like a symmetron ($Q_A<0$) \cite{Hinterbichler:2010es} or scalarisation \cite{Damour:1993hw} ($Q_A>0$). The effects of the coupling to the gluons will be analysed below.

The field at infinity oscillates as
\be
\varphi_\infty(t) = \varphi_0 \cos(m_\phi t)\,,
\ee
and must be regular at the centre of the ball of uniform density $\rho_A$,
mass $M_A$ and radius $R_A$.
These boundary conditions determine the field profile, which takes the form \cite{Hees:2018fpg}
\be
\varphi= \varphi_0 \cos(m_\phi t) f_\pm(r/R_A) \,.
\ee
The sign of the charge $Q_A$ selects either the function $f_+$ or $f_-$, with
\begin{eqnarray}
&&Q_A < 0 : \;\; x < 1 , \;\;\;  f_+(x) = \frac{1}{\cosh u} \frac{\sinh u x}{u x} \;\; \mbox{and for} \;\;
x>1 , \;\;\; f_+(x) = 1 - \frac{1}{x} \left( 1 - \frac{\tanh u}{u} \right) ,
\nonumber \\
&&
Q_A > 0 : \;\; x < 1 , \;\;\;  f_-(x) = \frac{1}{\cos u} \frac{\sin u x}{u x} \;\; \mbox{and for} \;\;
x>1 , \;\;\; f_-(x) = 1 - \frac{1}{x} \left( 1 - \frac{\tan u}{u} \right) ,
\nonumber \\
\end{eqnarray}
with
\be
u = \sqrt{ |Q_A| \frac{\rho_A R_A^2}{\Lambda_f^2 } }
= \sqrt{ |Q_A| \Phi_A \frac{6 m_{\rm Pl}^2}{\Lambda_f^2} } \,,
\label{eq:defscree}
\ee
where we introduced the gravitational potential at the surface of the object,
$\Phi_A = G_N M_A/R_A$.
In both cases, outside the object we have
\be
r > R_A : \;\;\; \varphi= \varphi_0 \left(1- s_A \frac{G_N M_A}{r}\right) \cos(m_\phi t)
= \varphi_\infty(t) - \frac{\beta_A(t)}{4\pi m_{\rm Pl}} \frac{M_A}{r} \,,
\label{screen}
\ee
where we introduced the couplings $s_A$ and $\beta_A(t)$, which are related by
\be
\beta_A(t)= \frac{s_A \varphi_\infty(t)}{2m_{\rm Pl}}\,.
\ee
In the unscreened regime, the couplings that give the amplitude of the fifth force outside of the
objects are
\be
\mbox{unscreened:} \;\;\; |u| \ll 1, \;\;\; s_A = \frac{2 Q_A m_{\rm Pl}^2}{\Lambda_f^2} \,, \;\;\;
\beta_A(t) = \frac{Q_A m_{\rm Pl} \varphi_\infty(t)}{\Lambda_f^2} = \alpha_A(\varphi(t))\,,
\ee
and the field is almost constant inside the object. This coupling is the same as the one of a point particle, i.e. no effect results from the finite size of the object.

On the other hand,
in the screened regime associated with large Newtonian potential $\Phi_A$,
the couplings are
\be
\mbox{screened:} \;\;\; |u| \gg 1, \;\;\; s_A = \frac{1}{\Phi_A} , \;\;\;
\beta_A(t) = \frac{\varphi_\infty(t)}{2 m_{\rm Pl} \Phi_A} ,
\ee
and the field shows a steep decay or fast oscillations inside the object. Notice that objects with large Newtonian potentials tend to be screened, a result reminiscent of screening mechanisms in modified gravity \cite{Brax:2012gr}. We will comment on this analogy below.

The coupling of the scalar to matter differs in the screened and unscreened cases. It is only species-dependent in the unscreened regime, through the charge $Q_A$. On the other hand, it is object-dependent in the screened regime through the dependence on the Newtonian potential $\Phi_A$. Objects with various Newton potentials have differing trajectories depending on the couplings $\beta_A\propto 1/\Phi_A$.

\subsection{Screened modified gravity}
The screening regime and the resulting screening mechanism for the dark scalar is
 similar to the symmetron screening mechanism, where the coupling to matter vanishes in regions of high density. Here and contrary to the symmetron case, the solution outside the object is the time-varying $\varphi_\infty$.  The screening criterion $|u| \gg 1$ is analogous to the one for  all non-derivative screening mechanisms \cite{Brax:2012gr}, i.e
\be
 \vert \beta_A^{\rm screened} \vert \leq \vert \beta_A^{\rm unscreened}\vert  : \;\;\;
\frac{\vert \varphi_\infty\vert }{2m_{\rm Pl} \Phi_A} \leq \frac{ m_{\rm Pl}\vert Q_A\vert  \vert \varphi_\infty\vert}{\Lambda_f^2} \,.
\ee
This corresponds to requiring that the effective coupling $\beta_A$ is less than the coupling $\alpha_A$ that an unscreened object such as a point particle would experience.
For static objects, the scalar force is proportional to the gradient of the scalar field.
Focusing on solar system tests where the test objects such as the Cassini satellite have  very small Newtonian potentials and behave like point particles in the scalar dark matter background,
screening may occur if the Earth or the Sun themselves are screened.
Thus, if the Earth is screened, the field gradient is suppressed and no deviation from General Relativity will take place in the vicinity of the Earth.
The Cassini bound \cite{Bertotti:2003rm} implies that
\be
\langle \beta_\oplus \alpha_S \rangle \le 2\times 10^{-5} \,,
\ee
where $\oplus$ denotes quantities evaluated for the Earth, and $S$ for the satellite, and  $Q_\oplus\sim 0.1$ for a model of the Earth made of silicon and iron, and
$Q_S\sim 0.1$ for a metallic satellite. We have taken the average over the rapid oscillations of the scalar field. This bound depends on the dark matter density locally $\rho_0= m^2_\phi \varphi_0^2/2$, which is of the order of $10^6$ times the cosmological matter density. This becomes the numerical constraint
\be
\frac{Q_S  \rho_0 }{m^2_\phi \Phi_\oplus \Lambda_f^2}\le 2\times 10^{-5}.
\ee
Using equation (\ref{eq:defscree}), we see that the Earth is screened provided
\be
\frac{\Lambda_f}{m_{\rm Pl}} \ll \sqrt{ {2Q_\oplus}{\Phi_\oplus}}\,,
\ee
where $\Phi_\oplus\simeq 10^{-9}$.
For the typical  H/He composition of the Sun we have $\vert Q_\odot\vert \simeq 0.15$ and $\Phi_\odot \sim 10^{-6}$, we see that the Sun is automatically screened if the Earth is screened.

\subsection{The E\"otv\"os parameter}

The contribution to the acceleration of an unscreened body $A$ in the field $\varphi$ due to the environment is given by
\be
\vec a_A^{\;\varphi} = -\frac{\beta_A(\varphi)}{m_{\rm Pl}}(\vec \nabla \varphi + \vec v_A \dot \varphi)\,,
\ee
where $\vec v_A$ is the non-relativistic velocity of body $A$.
Here we have
\be
\frac{\beta_A(\varphi)}{m_{\rm Pl}}= \frac{Q_A \varphi}{\Lambda_f^2}\,.
\ee
The acceleration of the test body $A$ due to the scalar field $\varphi$ generated by a distant
massive body $C$ is then given by
\be
\vec a_A^{\;\varphi} =Q_A \frac{\varphi_0^2}{\Lambda_f^2} \cos^2(m_\phi t)
\left( 1-s_C \frac{G_N M_C}{r} \right)
\left[ s_C \frac{G_N M_C}{r^3}\vec r - m_\phi \vec v_A \left(1-s_C \frac{G_N M_C}{r} \right)
\tan(m_\phi t) \right].
\ee
This is only the acceleration due to the scalar to which the gravitational acceleration should be added.

Let us now consider two test bodies $A$ and $B$ at the same location $r$ from a distant object $C$,
e.g. the two cylinders of the MICROSCOPE experiment aboard a satellite at 710 km from the Earth in a nearly circular orbit and falling in the terrestrial gravitational field \cite{Berge:2017ovy}. Then the difference between their accelerations towards the third object $C$ is given by
\begin{align}
\vec a_A^{\;\varphi} -\vec a_B^{\;\varphi} =&(Q_A-Q_B) \frac{\varphi_0^2}{\Lambda_f^2}
\cos^2(m_\phi t) \left(1-s_C \frac{G_N M_C}{r}\right)\nonumber\\
&\times \left[ s_C \frac{G_N M_C}{r^3}\vec r - m_\phi \vec v \left(1-s_C \frac{G_N M_C}{r}\right) \tan(m_\phi t) \right] ,
\end{align}
where we have taken that the centres of mass of the two objects coincide.
We see that the factor $(1-s_c G_N M_C/r)$  modulates the Newtonian acceleration and vanishes when $r=R_C$, the radius of object $C$, if object $C$ is screened.  Hence the violation of the equivalence principle are maximised for satellite experiments and minimised on the Earth, which must be screened for usual modified-gravity scenarios to pass terrestrial tests of gravity. There is an extra modulation when the objects move around $C$ with a common velocity $\vec v$.

We define the E\"otv\"os parameter
\be
\eta_{AB}=2 \left| \frac{\vec a_A -\vec a_B}{\vec a_A +\vec a_B} \right| \simeq
\left| \frac{\vec a_A^{\;\varphi} -\vec a_B^{\;\varphi}}{\vec a^N} \right| \,,
\ee
where $\vec a^N= - G_N M_C \vec r/r^3$ is the common Newtonian acceleration of the two bodies.
For quasi-circular orbits and on average this gives
\be
\eta_{AB}= \frac{\varphi_0^2}{2\Lambda_f^2}\vert Q_A-Q_B\vert s_C \left(1-s_C \frac{G_N M_C}{r}\right) \,,
\ee
where $r$ is the radius of the orbit.
If the Earth is screened, then for experiments such as MICROSCOPE we have $s_C=s_\oplus \simeq 1/\Phi_\oplus\simeq 10^9$. This implies that
\be
\eta_{AB}= \frac{\varphi_0^2}{\Lambda_f^2}\frac{\vert Q_A-Q_B\vert}{\Phi_\oplus}
\left(1- \frac{R_\oplus}{r}\right) \,.
\ee
Moreover in this experiment two cylinders of Platinum and Titanium alloys were used with $Q_{\rm Ti}-Q\sim -10^{-2}$
and $Q_{\rm Pt}-Q\sim -7.5\ 10^{-3}$
and $\vert Q_{\rm Ti}-Q_{\rm Pt}\vert\sim 2.9\ 10^{-3}$. As a result we get a bound from $\vert \eta_{\rm Pt-Ti}\vert \le 5 \ 10^{-15}$ \cite{MICROSCOPE:2022doy} on the amplitude
\be
\frac{\varphi_0}{\Lambda_f}\le  10^{-10} \,.
\ee
This can be used to put bounds on the parameters of the model as we can write
\be
\frac{\sqrt{\rho_0}}{m_\phi \Lambda_f}\le  10^{-10}\,,
\ee
which relates the cut-off of the theory, the mass of the dark matter field and the local density of dark matter. Fixing the local dark matter density  also implies
\be
\sqrt{m_\phi \Lambda_f} > 10^{-5} \; {\rm GeV} \,,
\label{bbi}
\ee
which correlates the suppression scale of the quadratic coupling to matter and the scalar mass.

\subsection{Constraints on the dilaton model parameters}

\subsubsection{Unscreened case}

For scalar masses below $1 \; {\rm eV}$, it is straightforward to check that the Earth and the Sun are both unscreened implying that the scalar field behaves like a nearly massless (when the range is large enough) field in the solar system with an effective coupling to matter
\be
\beta_A= \frac{Q_A m_{\rm Pl}\varphi_0}{\Lambda_f^2} \,.
\ee
Using $\varphi_0=\sqrt{2\rho_0}/m_\phi$, the effective coupling is of order
\be
\beta_A\simeq \frac{Q_A}{g\lambda} \frac{m_\phi m_{\rm Pl}\sqrt \rho_0}{8 \sqrt{2} v^2 \Lambda_c^2} \sim 10^{-24} \frac{1}{g\lambda} \frac{Q_A}{0.1} \frac{m_\phi}{1 \; {\rm eV}} \left( \frac{\Lambda_c}{1 \; {\rm TeV}} \right)^{-2} .
\ee
This small value of $\beta_A$ guarantees that the MICROSCOPE results are hardly affected by the scalar field.

Moreover, solar system tests are evaded when the coupling is smaller than the Cassini bound of order $10^{-5}$ for $\beta_A^2$. This is easily satisfied due to the very small value of the dark matter density and the smallness of the scalar mass. Therefore, even when the scalar is very light and the screening mechanism does not operate, the fact that the coupling to matter is proportional to $\varphi_0$ which is very small implies that the scalar is effectively decoupled from matter. The coupling becomes large enough to lead to possibly detectable effects only if locally the density of dark matter were to increase, for example due to the presence of a dark matter clump, significantly above the dark matter halo density \cite{Afach:2021pfd}. The investigation of this possibility is left for future work.

\subsubsection{Screened Case}
To see where in the dilaton parameter space screening might be relevant for terrestrial and satellite experiments, we start from the quadratic coupling
\be
\frac{1}{\Lambda_f^2}=-\frac{\partial_\phi^2 \mu^2}{2\mu^2} \,,
\ee
where we recall that the linear coupling vanishes as $\partial_\phi \mu=0$ at the minimum of the potential.
Explicitly we obtain that
\be
\frac{1}{\Lambda_f}=\frac{\pi^{3/2}}{4} \frac{M}{\sqrt \lambda  fv}\sqrt{ - \delta_{\rm min}  } \,,
\ee
or in terms of the scalar mass
\be
\frac{1}{\Lambda_f}=  \frac{1}{4\sqrt{g\lambda}}\frac{m_\phi}{v\Lambda_c} \,.
\ee
We get the bound from Eq.~(\ref{bbi})
\be
\sqrt{g\lambda} \Lambda_c >  10^{-13}~ {\rm GeV} \,.
\ee
This is always easily satisfied for theories with a cut-off scale larger than the electroweak scale.

As a result, the MICROSCOPE bound can always be respected due to the screening of the Earth if
\be
m_\phi > 1.9 \times 10^8 \frac{\sqrt{g\lambda}\Lambda_c}{m_{\rm Pl}} \; {\rm GeV} \,.
\ee
The lowest admissible cut-off scale from the particle physics point of view is $\Lambda_c \gtrsim 10$ TeV which implies that
\be
m_\phi > 1 \; {\rm keV} \,.
\ee
If the Earth were unscreened, the Sun could still be screened itself. This happens typically when
\be
m_\phi > 10 \; {\rm eV} \,,
\ee
as the Newtonian potential of the Sun is $10^3$ stronger than the Earth's
(but with a scalar charge that is five times larger). Screening of the Sun would guarantee that the Cassini experiment was insensitive to scalar interactions.
In conclusion, we find that screening could only occur for large scalar masses. Unfortunately, this is not allowed for scalar fields generating dark matter via the misalignment mechanism. If the scalar dark matter represented only a fraction of the dark matter density, the constraint on its mass would be relaxed and therefore one could envisage that the Earth could be screened. In all cases,  the dilaton studied here turns out to be invisible gravitationally. We now turn to other potential probes of the models under investigation.


\section{Further Phenomenology}
\subsection{Atomic Clocks}
Oscillating dark matter fields coupled to matter could lead to changes in atomic structure \cite{Arvanitaki:2014faa, Hees:2018fpg,Kennedy:2020bac}. In particular, tiny oscillations in the electron to proton mass ratio could be detected using atomic clocks. The variation of the atomic frequencies for various atomic transitions are sensitive to the coupling of the scalar field to particles such as the electrons and could probe very small couplings for very light scalars of masses less than $10^{-18}$ eV. In our model and in the background of the local dark matter density, the coupling of the scalar to matter particles is universal with a value\footnote{The coupling to fermions in \cite{Arvanitaki:2014faa} $d_i$ where $i$ labels the different fermions is universal in our model and equal to $\sqrt 2 \beta$. The same factor of $\sqrt 2 $ relates also the coupling to the QCD condensate and photons in the parameterisation of \cite{Damour:2010rm,Hees:2018fpg} and the couplings $(\beta_{\rm QCD},\beta_\gamma)$, see below. }
\be
\beta= \frac{m_{\rm Pl}\varphi_0}{\Lambda_f^2}\,,
\label{smcou}
\ee
of order $\beta \simeq 10^{-24} (\frac{m_\phi}{1\ \rm  eV})$ which is very much lower than the expected sensitivity of atomic clock experiments \cite{Hees:2018fpg}.

\subsection{Large Scale Structure}

The coupling of dark matter to baryons in cosmology could lead to an increase in the rate of growth for baryonic structures. Indeed and as long as the mass of the dark matter scalar is small enough, structures characterised by their wave number $k$ would be affected as long as $k/a\gtrsim m_\phi$, where $a$ is the scale factor of the Universe normalised to unity now. Gravity would be enhanced corresponding to a rescaling of Newton's constant by a factor
$(1+ 2 \beta^2(\rho))$ where the coupling is $\beta(\rho)= a^{-3/2} \beta$. As this coupling is valid from the electroweak scale time characterised by  a redshift $z_{\rm EW}$, where  $a^{-1}= 1+z$,  $z_{\rm EW}\simeq v/\Lambda_{\rm DE}\simeq 10^{14}$, we find that the smallness of the coupling (\ref{smcou}) cannot be compensated by the large redshift-dependent factor. This implies that no effects on the growth of structures is expected.

\subsection{Consequences of the coupling to bosons}
\label{sec:conse}

The  interactions between the light scalar and photons and gluons in equation (\ref{eq:bosons})  induce a dependence of the electromagnetic and the QCD couplings on the scalar field as
\begin{eqnarray}
&&
\frac{1}{e^2(\varphi)}= \frac{1}{e^2} + \alpha_F e^2  \frac{\varphi^2}{\Lambda^2_f}\,,\nonumber \\
&&
\frac{1}{g_3^2(\varphi)}= \frac{1}{g_3^2}+ \alpha_G g_3^2  \frac{\varphi^2}{\Lambda^2_f}\,.\nonumber \\
\end{eqnarray}
In the dark matter background this leads to linear couplings to the scalar field proportional to $\beta$. As a result both the fine structure constant and the QCD condensation scale, defined as the point where the QCD gauge coupling becomes large, become dependent on the scalar field. This  leads to contributions to the masses of the nucleons coming from the electromagnetic and gluonic energies \cite{Damour:1994zq,Brax:2006dc}. It turns out that the gluonic contribution dominates as the masses of nucleons are mostly due to the gluon condensate and we now concentrate on this effect.

QCD condensation takes place below the $c,b,t$ quark masses in terms of energy scale at a value around $250$ MeV. As a result, we only take into account the quadratic coupling of the scalar to the gluons when the heavy  quarks have been integrated out.
Writing the renormalisation group equation for the QCD coupling between the charm  decoupling scale $m_c$  and a lower scale $E$ we have
\be
\frac{4\pi}{g_3^2(E)}= \frac{4\pi}{g_3^2(\varphi)} -\frac{b_3}{2\pi}\ln \left(\frac{m_c}{E}\right)\,.
\ee
 The coefficient $b_3>0$ is the QCD beta function coefficient due to the gluons and the $u,d,s$ quarks . The QCD scale is such that $g_3(\Lambda_{\rm QCD})$ diverges leading to
\be
\Lambda_{\rm QCD}(\varphi)= m_c e^{-8\pi^2/b_3 g_3^2(\varphi)}\simeq \Lambda_{\rm QCD}\left(1- \frac{8\pi^2\alpha_G g_3^2}{b_3} \frac{\varphi^2}{\Lambda^2_f}\right)\,,
\ee
where in the last term $g_3$ is the QCD coupling at the energy scale $m_c$.
The quadratic dependence of the QCD scale on $\varphi^2$ can be tested using atomic clocks \cite{Hees:2018fpg} and places the constraint that for scalar masses $m_\phi \lesssim 10^{-18}$ eV  one must require that $\Lambda_f \gtrsim m_{\rm Pl}$. This is  easily achieved as $\Lambda_f \simeq \frac{v\Lambda_c}{m_\phi}$ where the scalar mass is lower than $10^{-18}$ eV  and the cut-off scale above the 10 TeV range.

In the dark matter background,
the dependence on the scalar variation $\delta \phi$ of $\Lambda_{\rm QCD}$ compared to the dark matter background $\varphi_0$ can be parameterised as
\be
\Lambda_{QCD}(\delta \varphi)= \Lambda_{QCD}\left(1+ \beta_{QCD} \frac{\delta\varphi}{m_{\rm Pl}}\right)\,,
\ee
where
\be
\beta_{\rm QCD}= - \frac{16\pi^2\alpha_G g_3^2}{b_3}\, \beta \,.
\ee
As this coupling is also very small and proportional to $\beta$, the conclusion that the scalar field hardly couples to matter and is therefore invisible in gravitational experiments remains\footnote{ The coupling $\alpha_A$ in the unscreened case is modified and is not given by (\ref{couling}) anymore \cite{Hees:2018fpg}. It becomes
$
\alpha_A= Q_A\beta
$
where  $Q_A= - \frac{16\pi^2\alpha_G g_3^2}{b_3} + (0.093+[Q'_A]_{\hat m}+ [Q'_A]_{ m_e}+[Q'_A]_{\delta m}) (1+ \frac{16\pi^2\alpha_G g_3^2}{b_3}) $.}.

Finally the dark matter  scalar can also decay to photons where the coupling is  induced at one loop. This can only happen when the scalar field gets a vev, which can occur  in the dark matter halo and also in the early Universe.
The decay rate to photons due to due fermions loops is given by
\be
\Gamma_{\phi\to \gamma\gamma}\simeq \beta_\gamma^2 \frac{m^3_\phi}{m_{\rm Pl}^2}\,,
\ee
where $\beta_\gamma= \alpha_F e^2 \frac{m_{\rm Pl}\varphi_0}{2\Lambda^2_f}$.
This is of order
\be
\Gamma_{\phi\to \gamma\gamma}\simeq e^4 \frac{m_\phi\rho_0}{\Lambda_f^4}\sim e^4 \frac{m_\phi^5 \rho_0}{v^4 \Lambda_c^4}\,.
\ee
This should be much smaller than $H_0 \sim \frac{\sqrt \rho_0}{m_{\rm Pl}}$, leading to a very weak bound on
\be
m_{\phi}\lesssim \left(\frac{v^4\Lambda_c^4}{e^4 m_{\rm Pl} \rho_0^{1/2}}\right)^{1/5}\,,
\ee
which is always satisfied easily for small $m_\phi\lesssim 1$ eV. Even in the early Universe where the constraint on a very slow decay rate is obtained by substituting $\rho_0\to \rho$, we find that the decay is essentially non-existent. This confirms that dark matter for these models is stable.

\section{Conclusion}
\label{sec:conclusion}

If we do not take into account the massive sector of neutrinos, the SM of particles and interactions only has one term with an explicit dimensional parameter. It is the $\mu$ term, which determines the vacuum expectation value of the Higgs field at low energies. In this work, we have studied the phenomenology associated with a dynamical $\mu$ term related to a new scalar degree of freedom. This field is coupled through the energy-momentum tensor of the matter content and can be identified with a dilaton associated with the conformal symmetry breaking of the theory in the matter sector only. We have discussed this framework by assuming a Higgs singlet and modeling the matter content with a unique fermion field. Interestingly, when the scalar sector of the model is stabilised at its fundamental state, the linear coupling of the dilaton to matter  disappears. This fact provides a very distinctive phenomenology for this new scalar degree of freedom, whose main coupling is quadratic. Another feature that we have explicitly discussed in  this work is the stability against radiative corrections of this model.

After analysing the main theoretical characteristics of this dilaton  model, we have studied its cosmological evolution.  The cosmological evolution leads the dilaton to lie close to the minimum of its potential and the corresponding oscillations are described by an harmonic approximation. In this limit, the energy-momentum tensor of the dilaton behaves as DM if its value can be averaged over many oscillations. In fact, this is what happens and we have found  explicitly that this dilaton can be  a candidate for  DM. As we have commented, the dilaton is coupled quadratically to  matter, so it is stable. In addition, the strength of its coupling is typically suppressed, which means that it does not thermalise for a broad range of the parameter space of the model.

Finally, we have studied the phenomelogical signatures of the model. The quadratic coupling  effectively couples  matter to the dilaton in a composition-dependent way. We have explored signals related to experiments measuring violations of the equivalence principle aboard satellites such as the MICROSCOPE experiment and future generation of tests related to this signature. We have found that the quadratic coupling provides naturally a type of screening mechanism similar to those studied for string motivated frameworks or symmetron models  \cite{Damour:1992kf,Damour:1993id,Cembranos:2009ds,Hinterbichler:2010es}. For masses below the electronvolt, the Sun and other planetary objects are not screened. Nevertheless, post-Newtonian parameter tests are easily fulfilled due to the the weakness of the strength of the effective linear coupling, even in the Solar system. This makes the dilaton invisible. Visibility would be granted if the dark matter density were locally much larger such as in a dark matter clump. Such clumps could result from the balance between the quantum pressure and the attraction due to the negative quartic interaction of the dilaton close to the electroweak minimum. The study of this possibility is left for future work.
\section*{Acknowledgements}

We would like to thank Raffaele d'Agnolo, Aur\'elien Hees, J\'er\'emie Quevillon and G\'eraldine Servant for interesting suggestions.
This work was partially supported by the MICINN (Ministerio de Ciencia e Innovación, Spain) project PID2019-107394GB-I00/AEI/10.13039/501100011033 (AEI/FEDER, UE) and the COST (European Cooperation in Science and Technology) Actions CosmicWISPers CA21106 and CosmoVerse CA2136. CB is supported by a Research Leadership Award from the Leverhulme Trust and by the STFC under grant ST/T000732/1. JARC acknowledges support by Institut Pascal at Université Paris-Saclay during the Paris-Saclay Astroparticle Symposium 2022, with the support of the P2IO Laboratory of Excellence (program “Investissements d’avenir” ANR-11-IDEX-0003-01 Paris-Saclay and ANR-10-LABX-0038), the P2I axis of the Graduate School of Physics of Université Paris-Saclay, as well as IJCLab, CEA, APPEC, IAS, OSUPS, and the IN2P3 master projet UCMN. CB is  supported by a Research Leadership Award from The Leverhulme Trust.

\appendix
\section{A possible model}
We consider a model with a massive field $X$ of large mass $m$ coupled to the Higgs field according to the Lagrangian
\be
{\cal L}= -g'(mX-u H^2)^2 -\lambda' X \bar \psi_H\psi_H \,,
\ee
where the fermions $\psi_H$ are charged under a gauge group which condenses in a similar way to QCD. At energies $E\ll m$, we can integrate out the field $X$ according to
\be
X= \frac{u H^2}{m} \,,
\ee
implying the Higgs field couples to the gauged fermions as
\be
{\cal L}= -\lambda'u \frac{H^2}{m} \bar \psi_H\psi_H \,.
\ee
Assuming that the fermions condense according to \be
\langle \bar \psi_H\psi_H\rangle = \Lambda^3_H e^{i\frac{\pi}{2}\frac{\phi}{f}}\,,
\ee
at a scale $\Lambda_H\ll m$, we have at low energy a potential term
\be
{\cal L}= -2\lambda'u  \frac{\Lambda_H^3}{m} H^2 \cos \frac{\pi}{2}\frac{\phi}{f}\,,
\ee
of the type used in the main text. We will assume that this transition happens before the end of inflation so that the $cos$ term is realised as soon as the field evolves at the end of inflation.

\section{Slow-rolling in the Radiative era}
\label{app:rad}

Initially, when the field is released at the end of inflation, the field starts moving slowly. Let us look for a simplified solution of the Klein-Gordon equation in this regime. We assume that in this slow roll regime we have
\be
H\dot \phi = \alpha V'\,,
\ee
where $\alpha$ is nearly constant. This is what would happen in the slow roll regime during inflation although here $\alpha$ will not be equal to $-1/3$. Using this ansatz and identifying  $m^2= V''$  we get
\be
\frac{\ddot\phi}{H\dot\phi}=\alpha \frac{m^2}{H^2}- \frac{\dot H}{H^2} \,,
\ee
and $\dot H= -\frac{3(1+\omega)}{2}  H^2$ where the equation of state in the radiation era is  $\omega=1/3$. Now using the Klein-Gordon equation $\ddot \phi +3H \dot \phi + V'=0$
we obtain
\be
\frac{m^2}{H^2}\alpha^2 + \frac{9+3\omega}{2}\alpha +1=0\,.
\ee
This implies that
\be
\alpha= \frac{H^2}{m^2}\left(-\frac{9+3\omega}{4}+\frac{1}{2}\sqrt{\left(\frac{9+3\omega}{2}\right)^2 -4 \frac{m^2}{H^2}}\right ).
\ee
Obviously, this is only valid when $m/H$ is small enough and varies very slowly.
In the slow roll regime and having initially  $m/H_{\rm inf}\ll 1$ implies that
\be
\alpha\simeq -\frac{2}{9+3 \omega}=-\frac{1}{5}\,.
\ee
When the equation of state is close to $-1$ during inflation we retrieve that slow-roll is realised with $\alpha=-1/3$.   During radiation domination the slow roll evolution is governed by
\be
\frac{dx}{d\tau}=-\frac{4\alpha}{\pi}  \tau \left(1-\epsilon- \sin \frac{\pi}{2} x\right) \,,
\label{ddd}
\ee
where we have denoted $\tau= m_0 t= \frac{m_0}{2H}$ and the potential reads
\be
V(\phi)= \left(\frac{2}{\pi}\right)^2 m_0^2 f^2 \left( \frac{\pi}{2}(1-\epsilon)(1-x) - \cos \frac{\pi}{2} x\right)\,.
\ee
This differential equation (\ref{ddd}) fails when $\tau={\cal O}(1)$.

The solution to (\ref{ddd}) with the initial condition $x(\tau=\tau_i)\simeq x_e$ where the slow roll regime starts is given by
\be
x(\tau)=\frac{4}{\pi} \arctan\left(\frac{1-\sqrt{2\epsilon-\epsilon^2}X}{1-\epsilon}\right) \,,
\ee
where
\be
X=\frac{a+b}{1-ab} \,,
\ee
and
\be
a={\tanh\left(\frac{\alpha}{2}\sqrt{2\epsilon-\epsilon^2}(\tau^2 -\tau_i^2)\right),\ b= \frac{1-(1-\epsilon) \tan\frac{\pi x_e}{4} }{\sqrt{2\epsilon-\epsilon^2}}} \,.
\ee
The field starts moving when $\tau\sim 0.5$ corresponding to $H\simeq m_0$.

\bibliography{clump}

\begin{thebibliography}{67}%
\makeatletter
\providecommand \@ifxundefined [1]{%
 \@ifx{#1\undefined}
}%
\providecommand \@ifnum [1]{%
 \ifnum #1\expandafter \@firstoftwo
 \else \expandafter \@secondoftwo
 \fi
}%
\providecommand \@ifx [1]{%
 \ifx #1\expandafter \@firstoftwo
 \else \expandafter \@secondoftwo
 \fi
}%
\providecommand \natexlab [1]{#1}%
\providecommand \enquote  [1]{``#1''}%
\providecommand \bibnamefont  [1]{#1}%
\providecommand \bibfnamefont [1]{#1}%
\providecommand \citenamefont [1]{#1}%
\providecommand \href@noop [0]{\@secondoftwo}%
\providecommand \href [0]{\begingroup \@sanitize@url \@href}%
\providecommand \@href[1]{\@@startlink{#1}\@@href}%
\providecommand \@@href[1]{\endgroup#1\@@endlink}%
\providecommand \@sanitize@url [0]{\catcode `\\12\catcode `\$12\catcode
  `\&12\catcode `\#12\catcode `\^12\catcode `\_12\catcode `\%12\relax}%
\providecommand \@@startlink[1]{}%
\providecommand \@@endlink[0]{}%
\providecommand \url  [0]{\begingroup\@sanitize@url \@url }%
\providecommand \@url [1]{\endgroup\@href {#1}{\urlprefix }}%
\providecommand \urlprefix  [0]{URL }%
\providecommand \Eprint [0]{\href }%
\providecommand \doibase [0]{http://dx.doi.org/}%
\providecommand \selectlanguage [0]{\@gobble}%
\providecommand \bibinfo  [0]{\@secondoftwo}%
\providecommand \bibfield  [0]{\@secondoftwo}%
\providecommand \translation [1]{[#1]}%
\providecommand \BibitemOpen [0]{}%
\providecommand \bibitemStop [0]{}%
\providecommand \bibitemNoStop [0]{.\EOS\space}%
\providecommand \EOS [0]{\spacefactor3000\relax}%
\providecommand \BibitemShut  [1]{\csname bibitem#1\endcsname}%
\let\auto@bib@innerbib\@empty
\bibitem [{\citenamefont {Ostriker}\ and\ \citenamefont
  {Steinhardt}(2003)}]{Ostriker:2003qj}%
  \BibitemOpen
  \bibfield  {author} {\bibinfo {author} {\bibfnamefont {J.~P.}\ \bibnamefont
  {Ostriker}}\ and\ \bibinfo {author} {\bibfnamefont {P.~J.}\ \bibnamefont
  {Steinhardt}},\ }\href {\doibase 10.1126/science.1085976} {\bibfield
  {journal} {\bibinfo  {journal} {Science}\ }\textbf {\bibinfo {volume}
  {300}},\ \bibinfo {pages} {1909} (\bibinfo {year} {2003})},\ \Eprint
  {http://arxiv.org/abs/astro-ph/0306402} {arXiv:astro-ph/0306402 [astro-ph]}
  \BibitemShut {NoStop}%
\bibitem [{\citenamefont {Weinberg}\ \emph {et~al.}(2015)\citenamefont
  {Weinberg}, \citenamefont {Bullock}, \citenamefont {Governato}, \citenamefont
  {Kuzio~de Naray},\ and\ \citenamefont {Peter}}]{Weinberg:2013aya}%
  \BibitemOpen
  \bibfield  {author} {\bibinfo {author} {\bibfnamefont {D.~H.}\ \bibnamefont
  {Weinberg}}, \bibinfo {author} {\bibfnamefont {J.~S.}\ \bibnamefont
  {Bullock}}, \bibinfo {author} {\bibfnamefont {F.}~\bibnamefont {Governato}},
  \bibinfo {author} {\bibfnamefont {R.}~\bibnamefont {Kuzio~de Naray}}, \ and\
  \bibinfo {author} {\bibfnamefont {A.~H.~G.}\ \bibnamefont {Peter}},\ }\href
  {\doibase 10.1073/pnas.1308716112} {\bibfield  {journal} {\bibinfo  {journal}
  {Proc. Nat. Acad. Sci.}\ }\textbf {\bibinfo {volume} {112}},\ \bibinfo
  {pages} {12249} (\bibinfo {year} {2015})},\ \Eprint
  {http://arxiv.org/abs/1306.0913} {arXiv:1306.0913 [astro-ph.CO]} \BibitemShut
  {NoStop}%
\bibitem [{\citenamefont {Pontzen}\ and\ \citenamefont
  {Governato}(2014)}]{Pontzen:2014lma}%
  \BibitemOpen
  \bibfield  {author} {\bibinfo {author} {\bibfnamefont {A.}~\bibnamefont
  {Pontzen}}\ and\ \bibinfo {author} {\bibfnamefont {F.}~\bibnamefont
  {Governato}},\ }\href {\doibase 10.1038/nature12953} {\bibfield  {journal}
  {\bibinfo  {journal} {Nature}\ }\textbf {\bibinfo {volume} {506}},\ \bibinfo
  {pages} {171} (\bibinfo {year} {2014})},\ \Eprint
  {http://arxiv.org/abs/1402.1764} {arXiv:1402.1764 [astro-ph.CO]} \BibitemShut
  {NoStop}%
\bibitem [{\citenamefont {Boylan-Kolchin}\ \emph {et~al.}(2011)\citenamefont
  {Boylan-Kolchin}, \citenamefont {Bullock},\ and\ \citenamefont
  {Kaplinghat}}]{BoylanKolchin:2011de}%
  \BibitemOpen
  \bibfield  {author} {\bibinfo {author} {\bibfnamefont {M.}~\bibnamefont
  {Boylan-Kolchin}}, \bibinfo {author} {\bibfnamefont {J.~S.}\ \bibnamefont
  {Bullock}}, \ and\ \bibinfo {author} {\bibfnamefont {M.}~\bibnamefont
  {Kaplinghat}},\ }\href {\doibase 10.1111/j.1745-3933.2011.01074.x} {\bibfield
   {journal} {\bibinfo  {journal} {Mon. Not. Roy. Astron. Soc.}\ }\textbf
  {\bibinfo {volume} {415}},\ \bibinfo {pages} {L40} (\bibinfo {year}
  {2011})},\ \Eprint {http://arxiv.org/abs/1103.0007} {arXiv:1103.0007
  [astro-ph.CO]} \BibitemShut {NoStop}%
\bibitem [{\citenamefont {Moore}\ \emph {et~al.}(1999)\citenamefont {Moore},
  \citenamefont {Ghigna}, \citenamefont {Governato}, \citenamefont {Lake},
  \citenamefont {Quinn}, \citenamefont {Stadel},\ and\ \citenamefont
  {Tozzi}}]{Moore:1999nt}%
  \BibitemOpen
  \bibfield  {author} {\bibinfo {author} {\bibfnamefont {B.}~\bibnamefont
  {Moore}}, \bibinfo {author} {\bibfnamefont {S.}~\bibnamefont {Ghigna}},
  \bibinfo {author} {\bibfnamefont {F.}~\bibnamefont {Governato}}, \bibinfo
  {author} {\bibfnamefont {G.}~\bibnamefont {Lake}}, \bibinfo {author}
  {\bibfnamefont {T.~R.}\ \bibnamefont {Quinn}}, \bibinfo {author}
  {\bibfnamefont {J.}~\bibnamefont {Stadel}}, \ and\ \bibinfo {author}
  {\bibfnamefont {P.}~\bibnamefont {Tozzi}},\ }\href {\doibase 10.1086/312287}
  {\bibfield  {journal} {\bibinfo  {journal} {Astrophys. J.}\ }\textbf
  {\bibinfo {volume} {524}},\ \bibinfo {pages} {L19} (\bibinfo {year}
  {1999})},\ \Eprint {http://arxiv.org/abs/astro-ph/9907411}
  {arXiv:astro-ph/9907411 [astro-ph]} \BibitemShut {NoStop}%
\bibitem [{\citenamefont {de~Blok}(2010)}]{deBlok:2009sp}%
  \BibitemOpen
  \bibfield  {author} {\bibinfo {author} {\bibfnamefont {W.~J.~G.}\
  \bibnamefont {de~Blok}},\ }\href {\doibase 10.1155/2010/789293} {\bibfield
  {journal} {\bibinfo  {journal} {Adv. Astron.}\ }\textbf {\bibinfo {volume}
  {2010}},\ \bibinfo {pages} {789293} (\bibinfo {year} {2010})},\ \Eprint
  {http://arxiv.org/abs/0910.3538} {arXiv:0910.3538 [astro-ph.CO]} \BibitemShut
  {NoStop}%
\bibitem [{\citenamefont {Cs\'aki}\ and\ \citenamefont
  {Tanedo}(2015)}]{Csaki:2015xpj}%
  \BibitemOpen
  \bibfield  {author} {\bibinfo {author} {\bibfnamefont {C.}~\bibnamefont
  {Cs\'aki}}\ and\ \bibinfo {author} {\bibfnamefont {P.}~\bibnamefont
  {Tanedo}},\ }in\ \href {\doibase 10.5170/CERN-2015-004.169} {\emph {\bibinfo
  {booktitle} {{2013 European School of High-Energy Physics}}}}\ (\bibinfo
  {year} {2015})\ pp.\ \bibinfo {pages} {169--268},\ \Eprint
  {http://arxiv.org/abs/1602.04228} {arXiv:1602.04228 [hep-ph]} \BibitemShut
  {NoStop}%
\bibitem [{\citenamefont {Feng}(2022)}]{Feng:2022rxt}%
  \BibitemOpen
  \bibfield  {author} {\bibinfo {author} {\bibfnamefont {J.~L.}\ \bibnamefont
  {Feng}},\ }in\ \href@noop {} {\emph {\bibinfo {booktitle} {{Les Houches
  summer school on Dark Matter}}}}\ (\bibinfo {year} {2022})\ \Eprint
  {http://arxiv.org/abs/2212.02479} {arXiv:2212.02479 [hep-ph]} \BibitemShut
  {NoStop}%
\bibitem [{\citenamefont {Banerjee}\ \emph {et~al.}(2019)\citenamefont
  {Banerjee}, \citenamefont {Kim},\ and\ \citenamefont
  {Perez}}]{Banerjee:2018xmn}%
  \BibitemOpen
  \bibfield  {author} {\bibinfo {author} {\bibfnamefont {A.}~\bibnamefont
  {Banerjee}}, \bibinfo {author} {\bibfnamefont {H.}~\bibnamefont {Kim}}, \
  and\ \bibinfo {author} {\bibfnamefont {G.}~\bibnamefont {Perez}},\ }\href
  {\doibase 10.1103/PhysRevD.100.115026} {\bibfield  {journal} {\bibinfo
  {journal} {Phys. Rev. D}\ }\textbf {\bibinfo {volume} {100}},\ \bibinfo
  {pages} {115026} (\bibinfo {year} {2019})},\ \Eprint
  {http://arxiv.org/abs/1810.01889} {arXiv:1810.01889 [hep-ph]} \BibitemShut
  {NoStop}%
\bibitem [{\citenamefont {Banerjee}\ \emph {et~al.}(2020)\citenamefont
  {Banerjee}, \citenamefont {Budker}, \citenamefont {Eby}, \citenamefont
  {Kim},\ and\ \citenamefont {Perez}}]{Banerjee:2019epw}%
  \BibitemOpen
  \bibfield  {author} {\bibinfo {author} {\bibfnamefont {A.}~\bibnamefont
  {Banerjee}}, \bibinfo {author} {\bibfnamefont {D.}~\bibnamefont {Budker}},
  \bibinfo {author} {\bibfnamefont {J.}~\bibnamefont {Eby}}, \bibinfo {author}
  {\bibfnamefont {H.}~\bibnamefont {Kim}}, \ and\ \bibinfo {author}
  {\bibfnamefont {G.}~\bibnamefont {Perez}},\ }\href {\doibase
  10.1038/s42005-019-0260-3} {\bibfield  {journal} {\bibinfo  {journal}
  {Commun. Phys.}\ }\textbf {\bibinfo {volume} {3}},\ \bibinfo {pages} {1}
  (\bibinfo {year} {2020})},\ \Eprint {http://arxiv.org/abs/1902.08212}
  {arXiv:1902.08212 [hep-ph]} \BibitemShut {NoStop}%
\bibitem [{\citenamefont {Banerjee}\ \emph {et~al.}(2021)\citenamefont
  {Banerjee}, \citenamefont {Madge}, \citenamefont {Perez}, \citenamefont
  {Ratzinger},\ and\ \citenamefont {Schwaller}}]{Banerjee:2021oeu}%
  \BibitemOpen
  \bibfield  {author} {\bibinfo {author} {\bibfnamefont {A.}~\bibnamefont
  {Banerjee}}, \bibinfo {author} {\bibfnamefont {E.}~\bibnamefont {Madge}},
  \bibinfo {author} {\bibfnamefont {G.}~\bibnamefont {Perez}}, \bibinfo
  {author} {\bibfnamefont {W.}~\bibnamefont {Ratzinger}}, \ and\ \bibinfo
  {author} {\bibfnamefont {P.}~\bibnamefont {Schwaller}},\ }\href {\doibase
  10.1103/PhysRevD.104.055026} {\bibfield  {journal} {\bibinfo  {journal}
  {Phys. Rev. D}\ }\textbf {\bibinfo {volume} {104}},\ \bibinfo {pages}
  {055026} (\bibinfo {year} {2021})},\ \Eprint
  {http://arxiv.org/abs/2105.12135} {arXiv:2105.12135 [hep-ph]} \BibitemShut
  {NoStop}%
\bibitem [{\citenamefont {Chatrchyan}\ and\ \citenamefont
  {Servant}(2022)}]{Chatrchyan:2022dpy}%
  \BibitemOpen
  \bibfield  {author} {\bibinfo {author} {\bibfnamefont {A.}~\bibnamefont
  {Chatrchyan}}\ and\ \bibinfo {author} {\bibfnamefont {G.}~\bibnamefont
  {Servant}},\ }\href@noop {} {\  (\bibinfo {year} {2022})},\ \Eprint
  {http://arxiv.org/abs/2211.15694} {arXiv:2211.15694 [hep-ph]} \BibitemShut
  {NoStop}%
\bibitem [{\citenamefont {Tito~D'Agnolo}\ and\ \citenamefont
  {Teresi}(2022)}]{TitoDAgnolo:2021pjo}%
  \BibitemOpen
  \bibfield  {author} {\bibinfo {author} {\bibfnamefont {R.}~\bibnamefont
  {Tito~D'Agnolo}}\ and\ \bibinfo {author} {\bibfnamefont {D.}~\bibnamefont
  {Teresi}},\ }\href {\doibase 10.1007/JHEP02(2022)023} {\bibfield  {journal}
  {\bibinfo  {journal} {JHEP}\ }\textbf {\bibinfo {volume} {02}},\ \bibinfo
  {pages} {023} (\bibinfo {year} {2022})},\ \Eprint
  {http://arxiv.org/abs/2109.13249} {arXiv:2109.13249 [hep-ph]} \BibitemShut
  {NoStop}%
\bibitem [{\citenamefont {Peccei}\ and\ \citenamefont
  {Quinn}(1977)}]{Peccei:1977hh}%
  \BibitemOpen
  \bibfield  {author} {\bibinfo {author} {\bibfnamefont {R.~D.}\ \bibnamefont
  {Peccei}}\ and\ \bibinfo {author} {\bibfnamefont {H.~R.}\ \bibnamefont
  {Quinn}},\ }\href {\doibase 10.1103/PhysRevLett.38.1440} {\bibfield
  {journal} {\bibinfo  {journal} {Phys. Rev. Lett.}\ }\textbf {\bibinfo
  {volume} {38}},\ \bibinfo {pages} {1440} (\bibinfo {year}
  {1977})}\BibitemShut {NoStop}%
\bibitem [{\citenamefont {Wilczek}(1978)}]{Wilczek:1977pj}%
  \BibitemOpen
  \bibfield  {author} {\bibinfo {author} {\bibfnamefont {F.}~\bibnamefont
  {Wilczek}},\ }\href {\doibase 10.1103/PhysRevLett.40.279} {\bibfield
  {journal} {\bibinfo  {journal} {Phys. Rev. Lett.}\ }\textbf {\bibinfo
  {volume} {40}},\ \bibinfo {pages} {279} (\bibinfo {year} {1978})}\BibitemShut
  {NoStop}%
\bibitem [{\citenamefont {Weinberg}(1978)}]{Weinberg:1977ma}%
  \BibitemOpen
  \bibfield  {author} {\bibinfo {author} {\bibfnamefont {S.}~\bibnamefont
  {Weinberg}},\ }\href {\doibase 10.1103/PhysRevLett.40.223} {\bibfield
  {journal} {\bibinfo  {journal} {Phys. Rev. Lett.}\ }\textbf {\bibinfo
  {volume} {40}},\ \bibinfo {pages} {223} (\bibinfo {year} {1978})}\BibitemShut
  {NoStop}%
\bibitem [{\citenamefont {Vysotsky}\ \emph {et~al.}(1978)\citenamefont
  {Vysotsky}, \citenamefont {Zeldovich}, \citenamefont {Khlopov},\ and\
  \citenamefont {Chechetkin}}]{Vysotsky:1978dc}%
  \BibitemOpen
  \bibfield  {author} {\bibinfo {author} {\bibfnamefont {M.~I.}\ \bibnamefont
  {Vysotsky}}, \bibinfo {author} {\bibfnamefont {Y.~B.}\ \bibnamefont
  {Zeldovich}}, \bibinfo {author} {\bibfnamefont {M.~Y.}\ \bibnamefont
  {Khlopov}}, \ and\ \bibinfo {author} {\bibfnamefont {V.~M.}\ \bibnamefont
  {Chechetkin}},\ }\href@noop {} {\bibfield  {journal} {\bibinfo  {journal}
  {Pisma Zh. Eksp. Teor. Fiz.}\ }\textbf {\bibinfo {volume} {27}},\ \bibinfo
  {pages} {533} (\bibinfo {year} {1978})}\BibitemShut {NoStop}%
\bibitem [{\citenamefont {Preskill}\ \emph {et~al.}(1983)\citenamefont
  {Preskill}, \citenamefont {Wise},\ and\ \citenamefont
  {Wilczek}}]{Preskill:1982cy}%
  \BibitemOpen
  \bibfield  {author} {\bibinfo {author} {\bibfnamefont {J.}~\bibnamefont
  {Preskill}}, \bibinfo {author} {\bibfnamefont {M.~B.}\ \bibnamefont {Wise}},
  \ and\ \bibinfo {author} {\bibfnamefont {F.}~\bibnamefont {Wilczek}},\ }\href
  {\doibase 10.1016/0370-2693(83)90637-8} {\bibfield  {journal} {\bibinfo
  {journal} {Phys. Lett. B}\ }\textbf {\bibinfo {volume} {120}},\ \bibinfo
  {pages} {127} (\bibinfo {year} {1983})}\BibitemShut {NoStop}%
\bibitem [{\citenamefont {Turner}\ \emph {et~al.}(1983)\citenamefont {Turner},
  \citenamefont {Wilczek},\ and\ \citenamefont {Zee}}]{Turner:1983sj}%
  \BibitemOpen
  \bibfield  {author} {\bibinfo {author} {\bibfnamefont {M.~S.}\ \bibnamefont
  {Turner}}, \bibinfo {author} {\bibfnamefont {F.}~\bibnamefont {Wilczek}}, \
  and\ \bibinfo {author} {\bibfnamefont {A.}~\bibnamefont {Zee}},\ }\href
  {\doibase 10.1016/0370-2693(83)91229-7} {\bibfield  {journal} {\bibinfo
  {journal} {Phys. Lett. B}\ }\textbf {\bibinfo {volume} {125}},\ \bibinfo
  {pages} {35} (\bibinfo {year} {1983})},\ \bibinfo {note} {[Erratum:
  Phys.Lett.B 125, 519 (1983)]}\BibitemShut {NoStop}%
\bibitem [{\citenamefont {Turner}(1983)}]{Turner:1983he}%
  \BibitemOpen
  \bibfield  {author} {\bibinfo {author} {\bibfnamefont {M.~S.}\ \bibnamefont
  {Turner}},\ }\href {\doibase 10.1103/PhysRevD.28.1243} {\bibfield  {journal}
  {\bibinfo  {journal} {Phys. Rev.}\ }\textbf {\bibinfo {volume} {D28}},\
  \bibinfo {pages} {1243} (\bibinfo {year} {1983})}\BibitemShut {NoStop}%
\bibitem [{\citenamefont {Ure\~na L\'opez}(2019)}]{Urena-Lopez:2019kud}%
  \BibitemOpen
  \bibfield  {author} {\bibinfo {author} {\bibfnamefont {L.~A.}\ \bibnamefont
  {Ure\~na L\'opez}},\ }\href {\doibase 10.3389/fspas.2019.00047} {\bibfield
  {journal} {\bibinfo  {journal} {Front. Astron. Space Sci.}\ }\textbf
  {\bibinfo {volume} {6}},\ \bibinfo {pages} {47} (\bibinfo {year}
  {2019})}\BibitemShut {NoStop}%
\bibitem [{\citenamefont {Sahni}\ and\ \citenamefont
  {Wang}(2000)}]{Sahni:1999qe}%
  \BibitemOpen
  \bibfield  {author} {\bibinfo {author} {\bibfnamefont {V.}~\bibnamefont
  {Sahni}}\ and\ \bibinfo {author} {\bibfnamefont {L.-M.}\ \bibnamefont
  {Wang}},\ }\href {\doibase 10.1103/PhysRevD.62.103517} {\bibfield  {journal}
  {\bibinfo  {journal} {Phys. Rev. D}\ }\textbf {\bibinfo {volume} {62}},\
  \bibinfo {pages} {103517} (\bibinfo {year} {2000})},\ \Eprint
  {http://arxiv.org/abs/astro-ph/9910097} {arXiv:astro-ph/9910097} \BibitemShut
  {NoStop}%
\bibitem [{\citenamefont {Johnson}\ and\ \citenamefont
  {Kamionkowski}(2008)}]{Johnson:2008se}%
  \BibitemOpen
  \bibfield  {author} {\bibinfo {author} {\bibfnamefont {M.~C.}\ \bibnamefont
  {Johnson}}\ and\ \bibinfo {author} {\bibfnamefont {M.}~\bibnamefont
  {Kamionkowski}},\ }\href {\doibase 10.1103/PhysRevD.78.063010} {\bibfield
  {journal} {\bibinfo  {journal} {Phys. Rev.}\ }\textbf {\bibinfo {volume}
  {D78}},\ \bibinfo {pages} {063010} (\bibinfo {year} {2008})},\ \Eprint
  {http://arxiv.org/abs/0805.1748} {arXiv:0805.1748 [astro-ph]} \BibitemShut
  {NoStop}%
\bibitem [{\citenamefont {Hu}\ \emph {et~al.}(2000)\citenamefont {Hu},
  \citenamefont {Barkana},\ and\ \citenamefont {Gruzinov}}]{Hu:2000ke}%
  \BibitemOpen
  \bibfield  {author} {\bibinfo {author} {\bibfnamefont {W.}~\bibnamefont
  {Hu}}, \bibinfo {author} {\bibfnamefont {R.}~\bibnamefont {Barkana}}, \ and\
  \bibinfo {author} {\bibfnamefont {A.}~\bibnamefont {Gruzinov}},\ }\href
  {\doibase 10.1103/PhysRevLett.85.1158} {\bibfield  {journal} {\bibinfo
  {journal} {Phys. Rev. Lett.}\ }\textbf {\bibinfo {volume} {85}},\ \bibinfo
  {pages} {1158} (\bibinfo {year} {2000})},\ \Eprint
  {http://arxiv.org/abs/astro-ph/0003365} {arXiv:astro-ph/0003365 [astro-ph]}
  \BibitemShut {NoStop}%
\bibitem [{\citenamefont {Hui}\ \emph {et~al.}(2017)\citenamefont {Hui},
  \citenamefont {Ostriker}, \citenamefont {Tremaine},\ and\ \citenamefont
  {Witten}}]{Hui:2016ltb}%
  \BibitemOpen
  \bibfield  {author} {\bibinfo {author} {\bibfnamefont {L.}~\bibnamefont
  {Hui}}, \bibinfo {author} {\bibfnamefont {J.~P.}\ \bibnamefont {Ostriker}},
  \bibinfo {author} {\bibfnamefont {S.}~\bibnamefont {Tremaine}}, \ and\
  \bibinfo {author} {\bibfnamefont {E.}~\bibnamefont {Witten}},\ }\href
  {\doibase 10.1103/PhysRevD.95.043541} {\bibfield  {journal} {\bibinfo
  {journal} {Phys. Rev. D}\ }\textbf {\bibinfo {volume} {95}},\ \bibinfo
  {pages} {043541} (\bibinfo {year} {2017})},\ \Eprint
  {http://arxiv.org/abs/1610.08297} {arXiv:1610.08297 [astro-ph.CO]}
  \BibitemShut {NoStop}%
\bibitem [{\citenamefont {Sakharov}\ and\ \citenamefont
  {Khlopov}(1994)}]{Sakharov:1994id}%
  \BibitemOpen
  \bibfield  {author} {\bibinfo {author} {\bibfnamefont {A.~S.}\ \bibnamefont
  {Sakharov}}\ and\ \bibinfo {author} {\bibfnamefont {M.~Y.}\ \bibnamefont
  {Khlopov}},\ }\href@noop {} {\bibfield  {journal} {\bibinfo  {journal} {Phys.
  Atom. Nucl.}\ }\textbf {\bibinfo {volume} {57}},\ \bibinfo {pages} {485}
  (\bibinfo {year} {1994})}\BibitemShut {NoStop}%
\bibitem [{\citenamefont {Sakharov}\ \emph {et~al.}(1996)\citenamefont
  {Sakharov}, \citenamefont {Sokoloff},\ and\ \citenamefont
  {Khlopov}}]{Sakharov:1996xg}%
  \BibitemOpen
  \bibfield  {author} {\bibinfo {author} {\bibfnamefont {A.~S.}\ \bibnamefont
  {Sakharov}}, \bibinfo {author} {\bibfnamefont {D.~D.}\ \bibnamefont
  {Sokoloff}}, \ and\ \bibinfo {author} {\bibfnamefont {M.~Y.}\ \bibnamefont
  {Khlopov}},\ }\href@noop {} {\bibfield  {journal} {\bibinfo  {journal} {Phys.
  Atom. Nucl.}\ }\textbf {\bibinfo {volume} {59}},\ \bibinfo {pages} {1005}
  (\bibinfo {year} {1996})}\BibitemShut {NoStop}%
\bibitem [{\citenamefont {Hwang}\ and\ \citenamefont
  {Noh}(2009)}]{Hwang:2009js}%
  \BibitemOpen
  \bibfield  {author} {\bibinfo {author} {\bibfnamefont {J.-c.}\ \bibnamefont
  {Hwang}}\ and\ \bibinfo {author} {\bibfnamefont {H.}~\bibnamefont {Noh}},\
  }\href {\doibase 10.1016/j.physletb.2009.08.031} {\bibfield  {journal}
  {\bibinfo  {journal} {Phys. Lett.}\ }\textbf {\bibinfo {volume} {B680}},\
  \bibinfo {pages} {1} (\bibinfo {year} {2009})},\ \Eprint
  {http://arxiv.org/abs/0902.4738} {arXiv:0902.4738 [astro-ph.CO]} \BibitemShut
  {NoStop}%
\bibitem [{\citenamefont {Park}\ \emph {et~al.}(2012)\citenamefont {Park},
  \citenamefont {Hwang},\ and\ \citenamefont {Noh}}]{Park:2012ru}%
  \BibitemOpen
  \bibfield  {author} {\bibinfo {author} {\bibfnamefont {C.-G.}\ \bibnamefont
  {Park}}, \bibinfo {author} {\bibfnamefont {J.-c.}\ \bibnamefont {Hwang}}, \
  and\ \bibinfo {author} {\bibfnamefont {H.}~\bibnamefont {Noh}},\ }\href
  {\doibase 10.1103/PhysRevD.86.083535} {\bibfield  {journal} {\bibinfo
  {journal} {Phys. Rev.}\ }\textbf {\bibinfo {volume} {D86}},\ \bibinfo {pages}
  {083535} (\bibinfo {year} {2012})},\ \Eprint {http://arxiv.org/abs/1207.3124}
  {arXiv:1207.3124 [astro-ph.CO]} \BibitemShut {NoStop}%
\bibitem [{\citenamefont {Hlozek}\ \emph {et~al.}(2015)\citenamefont {Hlozek},
  \citenamefont {Grin}, \citenamefont {Marsh},\ and\ \citenamefont
  {Ferreira}}]{Hlozek:2014lca}%
  \BibitemOpen
  \bibfield  {author} {\bibinfo {author} {\bibfnamefont {R.}~\bibnamefont
  {Hlozek}}, \bibinfo {author} {\bibfnamefont {D.}~\bibnamefont {Grin}},
  \bibinfo {author} {\bibfnamefont {D.~J.~E.}\ \bibnamefont {Marsh}}, \ and\
  \bibinfo {author} {\bibfnamefont {P.~G.}\ \bibnamefont {Ferreira}},\ }\href
  {\doibase 10.1103/PhysRevD.91.103512} {\bibfield  {journal} {\bibinfo
  {journal} {Phys. Rev.}\ }\textbf {\bibinfo {volume} {D91}},\ \bibinfo {pages}
  {103512} (\bibinfo {year} {2015})},\ \Eprint {http://arxiv.org/abs/1410.2896}
  {arXiv:1410.2896 [astro-ph.CO]} \BibitemShut {NoStop}%
\bibitem [{\citenamefont {Cembranos}\ \emph {et~al.}(2016)\citenamefont
  {Cembranos}, \citenamefont {Maroto},\ and\ \citenamefont
  {N{\'u}{\~n}ez~Jare{\~n}o}}]{Cembranos:2015oya}%
  \BibitemOpen
  \bibfield  {author} {\bibinfo {author} {\bibfnamefont {J.~A.~R.}\
  \bibnamefont {Cembranos}}, \bibinfo {author} {\bibfnamefont {A.~L.}\
  \bibnamefont {Maroto}}, \ and\ \bibinfo {author} {\bibfnamefont {S.~J.}\
  \bibnamefont {N{\'u}{\~n}ez~Jare{\~n}o}},\ }\href {\doibase
  10.1007/JHEP03(2016)013} {\bibfield  {journal} {\bibinfo  {journal} {JHEP}\
  }\textbf {\bibinfo {volume} {03}},\ \bibinfo {pages} {013} (\bibinfo {year}
  {2016})},\ \Eprint {http://arxiv.org/abs/1509.08819} {arXiv:1509.08819
  [astro-ph.CO]} \BibitemShut {NoStop}%
\bibitem [{\citenamefont {Cembranos}\ \emph {et~al.}(2017)\citenamefont
  {Cembranos}, \citenamefont {Maroto},\ and\ \citenamefont
  {N{\'u}{\~n}ez~Jare{\~n}o}}]{Cembranos:2016ugq}%
  \BibitemOpen
  \bibfield  {author} {\bibinfo {author} {\bibfnamefont {J.~A.~R.}\
  \bibnamefont {Cembranos}}, \bibinfo {author} {\bibfnamefont {A.~L.}\
  \bibnamefont {Maroto}}, \ and\ \bibinfo {author} {\bibfnamefont {S.~J.}\
  \bibnamefont {N{\'u}{\~n}ez~Jare{\~n}o}},\ }\href {\doibase
  10.1007/JHEP02(2017)064} {\bibfield  {journal} {\bibinfo  {journal} {JHEP}\
  }\textbf {\bibinfo {volume} {02}},\ \bibinfo {pages} {064} (\bibinfo {year}
  {2017})},\ \Eprint {http://arxiv.org/abs/1611.03793} {arXiv:1611.03793
  [astro-ph.CO]} \BibitemShut {NoStop}%
\bibitem [{\citenamefont {Schive}\ \emph {et~al.}(2014)\citenamefont {Schive},
  \citenamefont {Chiueh},\ and\ \citenamefont {Broadhurst}}]{Schive:2014dra}%
  \BibitemOpen
  \bibfield  {author} {\bibinfo {author} {\bibfnamefont {H.-Y.}\ \bibnamefont
  {Schive}}, \bibinfo {author} {\bibfnamefont {T.}~\bibnamefont {Chiueh}}, \
  and\ \bibinfo {author} {\bibfnamefont {T.}~\bibnamefont {Broadhurst}},\
  }\href {\doibase 10.1038/nphys2996} {\bibfield  {journal} {\bibinfo
  {journal} {Nature Phys.}\ }\textbf {\bibinfo {volume} {10}},\ \bibinfo
  {pages} {496} (\bibinfo {year} {2014})},\ \Eprint
  {http://arxiv.org/abs/1406.6586} {arXiv:1406.6586 [astro-ph.GA]} \BibitemShut
  {NoStop}%
\bibitem [{\citenamefont {Broadhurst}\ \emph {et~al.}(2018)\citenamefont
  {Broadhurst}, \citenamefont {Luu},\ and\ \citenamefont
  {Tye}}]{Broadhurst:2018fei}%
  \BibitemOpen
  \bibfield  {author} {\bibinfo {author} {\bibfnamefont {T.}~\bibnamefont
  {Broadhurst}}, \bibinfo {author} {\bibfnamefont {H.~N.}\ \bibnamefont {Luu}},
  \ and\ \bibinfo {author} {\bibfnamefont {S.~H.~H.}\ \bibnamefont {Tye}},\
  }\href@noop {} {\  (\bibinfo {year} {2018})},\ \Eprint
  {http://arxiv.org/abs/1811.03771} {arXiv:1811.03771 [astro-ph.GA]}
  \BibitemShut {NoStop}%
\bibitem [{\citenamefont {Cembranos}\ \emph {et~al.}(2005)\citenamefont
  {Cembranos}, \citenamefont {Feng}, \citenamefont {Rajaraman},\ and\
  \citenamefont {Takayama}}]{Cembranos:2005us}%
  \BibitemOpen
  \bibfield  {author} {\bibinfo {author} {\bibfnamefont {J.~A.~R.}\
  \bibnamefont {Cembranos}}, \bibinfo {author} {\bibfnamefont {J.~L.}\
  \bibnamefont {Feng}}, \bibinfo {author} {\bibfnamefont {A.}~\bibnamefont
  {Rajaraman}}, \ and\ \bibinfo {author} {\bibfnamefont {F.}~\bibnamefont
  {Takayama}},\ }\href {\doibase 10.1103/PhysRevLett.95.181301} {\bibfield
  {journal} {\bibinfo  {journal} {Phys. Rev. Lett.}\ }\textbf {\bibinfo
  {volume} {95}},\ \bibinfo {pages} {181301} (\bibinfo {year} {2005})},\
  \Eprint {http://arxiv.org/abs/hep-ph/0507150} {arXiv:hep-ph/0507150 [hep-ph]}
  \BibitemShut {NoStop}%
\bibitem [{\citenamefont {Cembranos}\ \emph {et~al.}(2018)\citenamefont
  {Cembranos}, \citenamefont {Maroto}, \citenamefont
  {N{\'u}{\~n}ez~Jare{\~n}o},\ and\ \citenamefont
  {Villarrubia-Rojo}}]{Cembranos:2018ulm}%
  \BibitemOpen
  \bibfield  {author} {\bibinfo {author} {\bibfnamefont {J.~A.~R.}\
  \bibnamefont {Cembranos}}, \bibinfo {author} {\bibfnamefont {A.~L.}\
  \bibnamefont {Maroto}}, \bibinfo {author} {\bibfnamefont {S.~J.}\
  \bibnamefont {N{\'u}{\~n}ez~Jare{\~n}o}}, \ and\ \bibinfo {author}
  {\bibfnamefont {H.}~\bibnamefont {Villarrubia-Rojo}},\ }\href {\doibase
  10.1007/JHEP08(2018)073} {\bibfield  {journal} {\bibinfo  {journal} {JHEP}\
  }\textbf {\bibinfo {volume} {08}},\ \bibinfo {pages} {073} (\bibinfo {year}
  {2018})},\ \Eprint {http://arxiv.org/abs/1805.08112} {arXiv:1805.08112
  [astro-ph.CO]} \BibitemShut {NoStop}%
\bibitem [{\citenamefont {Armengaud}\ \emph {et~al.}(2017)\citenamefont
  {Armengaud}, \citenamefont {Palanque-Delabrouille}, \citenamefont
  {Y{\`e}che}, \citenamefont {Marsh},\ and\ \citenamefont
  {Baur}}]{Armengaud:2017nkf}%
  \BibitemOpen
  \bibfield  {author} {\bibinfo {author} {\bibfnamefont {E.}~\bibnamefont
  {Armengaud}}, \bibinfo {author} {\bibfnamefont {N.}~\bibnamefont
  {Palanque-Delabrouille}}, \bibinfo {author} {\bibfnamefont {C.}~\bibnamefont
  {Y{\`e}che}}, \bibinfo {author} {\bibfnamefont {D.~J.~E.}\ \bibnamefont
  {Marsh}}, \ and\ \bibinfo {author} {\bibfnamefont {J.}~\bibnamefont {Baur}},\
  }\href {\doibase 10.1093/mnras/stx1870} {\bibfield  {journal} {\bibinfo
  {journal} {Mon. Not. Roy. Astron. Soc.}\ }\textbf {\bibinfo {volume} {471}},\
  \bibinfo {pages} {4606} (\bibinfo {year} {2017})},\ \Eprint
  {http://arxiv.org/abs/1703.09126} {arXiv:1703.09126 [astro-ph.CO]}
  \BibitemShut {NoStop}%
\bibitem [{\citenamefont {Brax}\ \emph
  {et~al.}(2019{\natexlab{a}})\citenamefont {Brax}, \citenamefont {Cembranos},\
  and\ \citenamefont {Valageas}}]{Brax:2019fzb}%
  \BibitemOpen
  \bibfield  {author} {\bibinfo {author} {\bibfnamefont {P.}~\bibnamefont
  {Brax}}, \bibinfo {author} {\bibfnamefont {J.~A.~R.}\ \bibnamefont
  {Cembranos}}, \ and\ \bibinfo {author} {\bibfnamefont {P.}~\bibnamefont
  {Valageas}},\ }\href {\doibase 10.1103/PhysRevD.100.023526} {\bibfield
  {journal} {\bibinfo  {journal} {Phys. Rev.}\ }\textbf {\bibinfo {volume}
  {D100}},\ \bibinfo {pages} {023526} (\bibinfo {year} {2019}{\natexlab{a}})},\
  \Eprint {http://arxiv.org/abs/1906.00730} {arXiv:1906.00730 [astro-ph.CO]}
  \BibitemShut {NoStop}%
\bibitem [{\citenamefont {Brax}\ \emph
  {et~al.}(2019{\natexlab{b}})\citenamefont {Brax}, \citenamefont {Valageas},\
  and\ \citenamefont {Cembranos}}]{Brax:2019npi}%
  \BibitemOpen
  \bibfield  {author} {\bibinfo {author} {\bibfnamefont {P.}~\bibnamefont
  {Brax}}, \bibinfo {author} {\bibfnamefont {P.}~\bibnamefont {Valageas}}, \
  and\ \bibinfo {author} {\bibfnamefont {J.~A.~R.}\ \bibnamefont {Cembranos}},\
  }\href@noop {} {\  (\bibinfo {year} {2019}{\natexlab{b}})},\ \Eprint
  {http://arxiv.org/abs/1909.02614} {arXiv:1909.02614 [astro-ph.CO]}
  \BibitemShut {NoStop}%
\bibitem [{\citenamefont {Brax}\ \emph
  {et~al.}(2020{\natexlab{a}})\citenamefont {Brax}, \citenamefont {Cembranos},\
  and\ \citenamefont {Valageas}}]{Brax:2020tuk}%
  \BibitemOpen
  \bibfield  {author} {\bibinfo {author} {\bibfnamefont {P.}~\bibnamefont
  {Brax}}, \bibinfo {author} {\bibfnamefont {J.~A.~R.}\ \bibnamefont
  {Cembranos}}, \ and\ \bibinfo {author} {\bibfnamefont {P.}~\bibnamefont
  {Valageas}},\ }\href {\doibase 10.1103/PhysRevD.101.063510} {\bibfield
  {journal} {\bibinfo  {journal} {Phys. Rev. D}\ }\textbf {\bibinfo {volume}
  {101}},\ \bibinfo {pages} {063510} (\bibinfo {year} {2020}{\natexlab{a}})},\
  \Eprint {http://arxiv.org/abs/2001.06873} {arXiv:2001.06873 [astro-ph.CO]}
  \BibitemShut {NoStop}%
\bibitem [{\citenamefont {Fonseca}\ \emph {et~al.}(2020)\citenamefont
  {Fonseca}, \citenamefont {Morgante}, \citenamefont {Sato},\ and\
  \citenamefont {Servant}}]{Fonseca:2019lmc}%
  \BibitemOpen
  \bibfield  {author} {\bibinfo {author} {\bibfnamefont {N.}~\bibnamefont
  {Fonseca}}, \bibinfo {author} {\bibfnamefont {E.}~\bibnamefont {Morgante}},
  \bibinfo {author} {\bibfnamefont {R.}~\bibnamefont {Sato}}, \ and\ \bibinfo
  {author} {\bibfnamefont {G.}~\bibnamefont {Servant}},\ }\href {\doibase
  10.1007/JHEP05(2020)080} {\bibfield  {journal} {\bibinfo  {journal} {JHEP}\
  }\textbf {\bibinfo {volume} {05}},\ \bibinfo {pages} {080} (\bibinfo {year}
  {2020})},\ \bibinfo {note} {[Erratum: JHEP 01, 012 (2021)]},\ \Eprint
  {http://arxiv.org/abs/1911.08473} {arXiv:1911.08473 [hep-ph]} \BibitemShut
  {NoStop}%
\bibitem [{\citenamefont {Goldberger}\ \emph {et~al.}(2008)\citenamefont
  {Goldberger}, \citenamefont {Grinstein},\ and\ \citenamefont
  {Skiba}}]{Goldberger:2007zk}%
  \BibitemOpen
  \bibfield  {author} {\bibinfo {author} {\bibfnamefont {W.~D.}\ \bibnamefont
  {Goldberger}}, \bibinfo {author} {\bibfnamefont {B.}~\bibnamefont
  {Grinstein}}, \ and\ \bibinfo {author} {\bibfnamefont {W.}~\bibnamefont
  {Skiba}},\ }\href {\doibase 10.1103/PhysRevLett.100.111802} {\bibfield
  {journal} {\bibinfo  {journal} {Phys. Rev. Lett.}\ }\textbf {\bibinfo
  {volume} {100}},\ \bibinfo {pages} {111802} (\bibinfo {year} {2008})},\
  \Eprint {http://arxiv.org/abs/0708.1463} {arXiv:0708.1463 [hep-ph]}
  \BibitemShut {NoStop}%
\bibitem [{\citenamefont {Damour}\ and\ \citenamefont
  {Nordtvedt}(1993{\natexlab{a}})}]{Damour:1992kf}%
  \BibitemOpen
  \bibfield  {author} {\bibinfo {author} {\bibfnamefont {T.}~\bibnamefont
  {Damour}}\ and\ \bibinfo {author} {\bibfnamefont {K.}~\bibnamefont
  {Nordtvedt}},\ }\href {\doibase 10.1103/PhysRevLett.70.2217} {\bibfield
  {journal} {\bibinfo  {journal} {Phys. Rev. Lett.}\ }\textbf {\bibinfo
  {volume} {70}},\ \bibinfo {pages} {2217} (\bibinfo {year}
  {1993}{\natexlab{a}})}\BibitemShut {NoStop}%
\bibitem [{\citenamefont {Damour}\ and\ \citenamefont
  {Nordtvedt}(1993{\natexlab{b}})}]{Damour:1993id}%
  \BibitemOpen
  \bibfield  {author} {\bibinfo {author} {\bibfnamefont {T.}~\bibnamefont
  {Damour}}\ and\ \bibinfo {author} {\bibfnamefont {K.}~\bibnamefont
  {Nordtvedt}},\ }\href {\doibase 10.1103/PhysRevD.48.3436} {\bibfield
  {journal} {\bibinfo  {journal} {Phys. Rev. D}\ }\textbf {\bibinfo {volume}
  {48}},\ \bibinfo {pages} {3436} (\bibinfo {year}
  {1993}{\natexlab{b}})}\BibitemShut {NoStop}%
\bibitem [{\citenamefont {Cembranos}\ \emph {et~al.}(2009)\citenamefont
  {Cembranos}, \citenamefont {Olive}, \citenamefont {Peloso},\ and\
  \citenamefont {Uzan}}]{Cembranos:2009ds}%
  \BibitemOpen
  \bibfield  {author} {\bibinfo {author} {\bibfnamefont {J.~A.~R.}\
  \bibnamefont {Cembranos}}, \bibinfo {author} {\bibfnamefont {K.~A.}\
  \bibnamefont {Olive}}, \bibinfo {author} {\bibfnamefont {M.}~\bibnamefont
  {Peloso}}, \ and\ \bibinfo {author} {\bibfnamefont {J.-P.}\ \bibnamefont
  {Uzan}},\ }\href {\doibase 10.1088/1475-7516/2009/07/025} {\bibfield
  {journal} {\bibinfo  {journal} {JCAP}\ }\textbf {\bibinfo {volume} {07}},\
  \bibinfo {pages} {025} (\bibinfo {year} {2009})},\ \Eprint
  {http://arxiv.org/abs/0905.1989} {arXiv:0905.1989 [astro-ph.CO]} \BibitemShut
  {NoStop}%
\bibitem [{\citenamefont {Hinterbichler}\ and\ \citenamefont
  {Khoury}(2010)}]{Hinterbichler:2010es}%
  \BibitemOpen
  \bibfield  {author} {\bibinfo {author} {\bibfnamefont {K.}~\bibnamefont
  {Hinterbichler}}\ and\ \bibinfo {author} {\bibfnamefont {J.}~\bibnamefont
  {Khoury}},\ }\href {\doibase 10.1103/PhysRevLett.104.231301} {\bibfield
  {journal} {\bibinfo  {journal} {Phys. Rev. Lett.}\ }\textbf {\bibinfo
  {volume} {104}},\ \bibinfo {pages} {231301} (\bibinfo {year} {2010})},\
  \Eprint {http://arxiv.org/abs/1001.4525} {arXiv:1001.4525 [hep-th]}
  \BibitemShut {NoStop}%
\bibitem [{\citenamefont {Damour}\ and\ \citenamefont
  {Esposito-Farese}(1993)}]{Damour:1993hw}%
  \BibitemOpen
  \bibfield  {author} {\bibinfo {author} {\bibfnamefont {T.}~\bibnamefont
  {Damour}}\ and\ \bibinfo {author} {\bibfnamefont {G.}~\bibnamefont
  {Esposito-Farese}},\ }\href {\doibase 10.1103/PhysRevLett.70.2220} {\bibfield
   {journal} {\bibinfo  {journal} {Phys. Rev. Lett.}\ }\textbf {\bibinfo
  {volume} {70}},\ \bibinfo {pages} {2220} (\bibinfo {year}
  {1993})}\BibitemShut {NoStop}%
\bibitem [{\citenamefont {Hees}\ \emph {et~al.}(2018)\citenamefont {Hees},
  \citenamefont {Minazzoli}, \citenamefont {Savalle}, \citenamefont {Stadnik},\
  and\ \citenamefont {Wolf}}]{Hees:2018fpg}%
  \BibitemOpen
  \bibfield  {author} {\bibinfo {author} {\bibfnamefont {A.}~\bibnamefont
  {Hees}}, \bibinfo {author} {\bibfnamefont {O.}~\bibnamefont {Minazzoli}},
  \bibinfo {author} {\bibfnamefont {E.}~\bibnamefont {Savalle}}, \bibinfo
  {author} {\bibfnamefont {Y.~V.}\ \bibnamefont {Stadnik}}, \ and\ \bibinfo
  {author} {\bibfnamefont {P.}~\bibnamefont {Wolf}},\ }\href {\doibase
  10.1103/PhysRevD.98.064051} {\bibfield  {journal} {\bibinfo  {journal} {Phys.
  Rev. D}\ }\textbf {\bibinfo {volume} {98}},\ \bibinfo {pages} {064051}
  (\bibinfo {year} {2018})},\ \Eprint {http://arxiv.org/abs/1807.04512}
  {arXiv:1807.04512 [gr-qc]} \BibitemShut {NoStop}%
\bibitem [{\citenamefont {Banerjee}\ \emph {et~al.}(2022)\citenamefont
  {Banerjee}, \citenamefont {Perez}, \citenamefont {Safronova}, \citenamefont
  {Savoray},\ and\ \citenamefont {Shalit}}]{Banerjee:2022sqg}%
  \BibitemOpen
  \bibfield  {author} {\bibinfo {author} {\bibfnamefont {A.}~\bibnamefont
  {Banerjee}}, \bibinfo {author} {\bibfnamefont {G.}~\bibnamefont {Perez}},
  \bibinfo {author} {\bibfnamefont {M.}~\bibnamefont {Safronova}}, \bibinfo
  {author} {\bibfnamefont {I.}~\bibnamefont {Savoray}}, \ and\ \bibinfo
  {author} {\bibfnamefont {A.}~\bibnamefont {Shalit}},\ }\href@noop {} {\
  (\bibinfo {year} {2022})},\ \Eprint {http://arxiv.org/abs/2211.05174}
  {arXiv:2211.05174 [hep-ph]} \BibitemShut {NoStop}%
\bibitem [{\citenamefont {Khoury}\ and\ \citenamefont
  {Weltman}(2004)}]{Khoury:2003rn}%
  \BibitemOpen
  \bibfield  {author} {\bibinfo {author} {\bibfnamefont {J.}~\bibnamefont
  {Khoury}}\ and\ \bibinfo {author} {\bibfnamefont {A.}~\bibnamefont
  {Weltman}},\ }\href {\doibase 10.1103/PhysRevD.69.044026} {\bibfield
  {journal} {\bibinfo  {journal} {Phys. Rev. D}\ }\textbf {\bibinfo {volume}
  {69}},\ \bibinfo {pages} {044026} (\bibinfo {year} {2004})},\ \Eprint
  {http://arxiv.org/abs/astro-ph/0309411} {arXiv:astro-ph/0309411} \BibitemShut
  {NoStop}%
\bibitem [{\citenamefont {Brax}\ and\ \citenamefont
  {Burrage}(2021)}]{Brax:2021rwk}%
  \BibitemOpen
  \bibfield  {author} {\bibinfo {author} {\bibfnamefont {P.}~\bibnamefont
  {Brax}}\ and\ \bibinfo {author} {\bibfnamefont {C.}~\bibnamefont {Burrage}},\
  }\href {\doibase 10.1103/PhysRevD.104.015011} {\bibfield  {journal} {\bibinfo
   {journal} {Phys. Rev. D}\ }\textbf {\bibinfo {volume} {104}},\ \bibinfo
  {pages} {015011} (\bibinfo {year} {2021})},\ \Eprint
  {http://arxiv.org/abs/2101.10693} {arXiv:2101.10693 [hep-ph]} \BibitemShut
  {NoStop}%
\bibitem [{\citenamefont {Coleman}\ and\ \citenamefont
  {Weinberg}(1973)}]{Coleman:1973jx}%
  \BibitemOpen
  \bibfield  {author} {\bibinfo {author} {\bibfnamefont {S.~R.}\ \bibnamefont
  {Coleman}}\ and\ \bibinfo {author} {\bibfnamefont {E.~J.}\ \bibnamefont
  {Weinberg}},\ }\href {\doibase 10.1103/PhysRevD.7.1888} {\bibfield  {journal}
  {\bibinfo  {journal} {Phys. Rev. D}\ }\textbf {\bibinfo {volume} {7}},\
  \bibinfo {pages} {1888} (\bibinfo {year} {1973})}\BibitemShut {NoStop}%
\bibitem [{\citenamefont {Brax}\ \emph {et~al.}(2011)\citenamefont {Brax},
  \citenamefont {Burrage}, \citenamefont {Davis}, \citenamefont {Seery},\ and\
  \citenamefont {Weltman}}]{Brax:2010uq}%
  \BibitemOpen
  \bibfield  {author} {\bibinfo {author} {\bibfnamefont {P.}~\bibnamefont
  {Brax}}, \bibinfo {author} {\bibfnamefont {C.}~\bibnamefont {Burrage}},
  \bibinfo {author} {\bibfnamefont {A.-C.}\ \bibnamefont {Davis}}, \bibinfo
  {author} {\bibfnamefont {D.}~\bibnamefont {Seery}}, \ and\ \bibinfo {author}
  {\bibfnamefont {A.}~\bibnamefont {Weltman}},\ }\href {\doibase
  10.1016/j.physletb.2011.03.047} {\bibfield  {journal} {\bibinfo  {journal}
  {Phys. Lett. B}\ }\textbf {\bibinfo {volume} {699}},\ \bibinfo {pages} {5}
  (\bibinfo {year} {2011})},\ \Eprint {http://arxiv.org/abs/1010.4536}
  {arXiv:1010.4536 [hep-th]} \BibitemShut {NoStop}%
\bibitem [{\citenamefont {Abbott}\ and\ \citenamefont
  {Sikivie}(1983)}]{Abbott:1982af}%
  \BibitemOpen
  \bibfield  {author} {\bibinfo {author} {\bibfnamefont {L.~F.}\ \bibnamefont
  {Abbott}}\ and\ \bibinfo {author} {\bibfnamefont {P.}~\bibnamefont
  {Sikivie}},\ }\href {\doibase 10.1016/0370-2693(83)90638-X} {\bibfield
  {journal} {\bibinfo  {journal} {Phys. Lett. B}\ }\textbf {\bibinfo {volume}
  {120}},\ \bibinfo {pages} {133} (\bibinfo {year} {1983})}\BibitemShut
  {NoStop}%
\bibitem [{\citenamefont {Dine}\ and\ \citenamefont
  {Fischler}(1983)}]{Dine:1982ah}%
  \BibitemOpen
  \bibfield  {author} {\bibinfo {author} {\bibfnamefont {M.}~\bibnamefont
  {Dine}}\ and\ \bibinfo {author} {\bibfnamefont {W.}~\bibnamefont
  {Fischler}},\ }\href {\doibase 10.1016/0370-2693(83)90639-1} {\bibfield
  {journal} {\bibinfo  {journal} {Phys. Lett. B}\ }\textbf {\bibinfo {volume}
  {120}},\ \bibinfo {pages} {137} (\bibinfo {year} {1983})}\BibitemShut
  {NoStop}%
\bibitem [{\citenamefont {Marsh}(2016)}]{Marsh:2015xka}%
  \BibitemOpen
  \bibfield  {author} {\bibinfo {author} {\bibfnamefont {D.~J.~E.}\
  \bibnamefont {Marsh}},\ }\href {\doibase 10.1016/j.physrep.2016.06.005}
  {\bibfield  {journal} {\bibinfo  {journal} {Phys. Rept.}\ }\textbf {\bibinfo
  {volume} {643}},\ \bibinfo {pages} {1} (\bibinfo {year} {2016})},\ \Eprint
  {http://arxiv.org/abs/1510.07633} {arXiv:1510.07633 [astro-ph.CO]}
  \BibitemShut {NoStop}%
\bibitem [{\citenamefont {Brax}\ \emph
  {et~al.}(2020{\natexlab{b}})\citenamefont {Brax}, \citenamefont {Cembranos},\
  and\ \citenamefont {Valageas}}]{Brax:2020oye}%
  \BibitemOpen
  \bibfield  {author} {\bibinfo {author} {\bibfnamefont {P.}~\bibnamefont
  {Brax}}, \bibinfo {author} {\bibfnamefont {J.~A.~R.}\ \bibnamefont
  {Cembranos}}, \ and\ \bibinfo {author} {\bibfnamefont {P.}~\bibnamefont
  {Valageas}},\ }\href {\doibase 10.1103/PhysRevD.102.083012} {\bibfield
  {journal} {\bibinfo  {journal} {Phys. Rev. D}\ }\textbf {\bibinfo {volume}
  {102}},\ \bibinfo {pages} {083012} (\bibinfo {year} {2020}{\natexlab{b}})},\
  \Eprint {http://arxiv.org/abs/2007.04638} {arXiv:2007.04638 [astro-ph.CO]}
  \BibitemShut {NoStop}%
\bibitem [{\citenamefont {Brax}\ \emph {et~al.}(2012)\citenamefont {Brax},
  \citenamefont {Davis}, \citenamefont {Li},\ and\ \citenamefont
  {Winther}}]{Brax:2012gr}%
  \BibitemOpen
  \bibfield  {author} {\bibinfo {author} {\bibfnamefont {P.}~\bibnamefont
  {Brax}}, \bibinfo {author} {\bibfnamefont {A.-C.}\ \bibnamefont {Davis}},
  \bibinfo {author} {\bibfnamefont {B.}~\bibnamefont {Li}}, \ and\ \bibinfo
  {author} {\bibfnamefont {H.~A.}\ \bibnamefont {Winther}},\ }\href {\doibase
  10.1103/PhysRevD.86.044015} {\bibfield  {journal} {\bibinfo  {journal} {Phys.
  Rev. D}\ }\textbf {\bibinfo {volume} {86}},\ \bibinfo {pages} {044015}
  (\bibinfo {year} {2012})},\ \Eprint {http://arxiv.org/abs/1203.4812}
  {arXiv:1203.4812 [astro-ph.CO]} \BibitemShut {NoStop}%
\bibitem [{\citenamefont {Bertotti}\ \emph {et~al.}(2003)\citenamefont
  {Bertotti}, \citenamefont {Iess},\ and\ \citenamefont
  {Tortora}}]{Bertotti:2003rm}%
  \BibitemOpen
  \bibfield  {author} {\bibinfo {author} {\bibfnamefont {B.}~\bibnamefont
  {Bertotti}}, \bibinfo {author} {\bibfnamefont {L.}~\bibnamefont {Iess}}, \
  and\ \bibinfo {author} {\bibfnamefont {P.}~\bibnamefont {Tortora}},\ }\href
  {\doibase 10.1038/nature01997} {\bibfield  {journal} {\bibinfo  {journal}
  {Nature}\ }\textbf {\bibinfo {volume} {425}},\ \bibinfo {pages} {374}
  (\bibinfo {year} {2003})}\BibitemShut {NoStop}%
\bibitem [{\citenamefont {Berg\'e}\ \emph {et~al.}(2018)\citenamefont
  {Berg\'e}, \citenamefont {Brax}, \citenamefont {M\'etris}, \citenamefont
  {Pernot-Borr\`as}, \citenamefont {Touboul},\ and\ \citenamefont
  {Uzan}}]{Berge:2017ovy}%
  \BibitemOpen
  \bibfield  {author} {\bibinfo {author} {\bibfnamefont {J.}~\bibnamefont
  {Berg\'e}}, \bibinfo {author} {\bibfnamefont {P.}~\bibnamefont {Brax}},
  \bibinfo {author} {\bibfnamefont {G.}~\bibnamefont {M\'etris}}, \bibinfo
  {author} {\bibfnamefont {M.}~\bibnamefont {Pernot-Borr\`as}}, \bibinfo
  {author} {\bibfnamefont {P.}~\bibnamefont {Touboul}}, \ and\ \bibinfo
  {author} {\bibfnamefont {J.-P.}\ \bibnamefont {Uzan}},\ }\href {\doibase
  10.1103/PhysRevLett.120.141101} {\bibfield  {journal} {\bibinfo  {journal}
  {Phys. Rev. Lett.}\ }\textbf {\bibinfo {volume} {120}},\ \bibinfo {pages}
  {141101} (\bibinfo {year} {2018})},\ \Eprint
  {http://arxiv.org/abs/1712.00483} {arXiv:1712.00483 [gr-qc]} \BibitemShut
  {NoStop}%
\bibitem [{\citenamefont {Touboul}\ \emph {et~al.}(2022)\citenamefont {Touboul}
  \emph {et~al.}}]{MICROSCOPE:2022doy}%
  \BibitemOpen
  \bibfield  {author} {\bibinfo {author} {\bibfnamefont {P.}~\bibnamefont
  {Touboul}} \emph {et~al.} (\bibinfo {collaboration} {MICROSCOPE}),\ }\href
  {\doibase 10.1103/PhysRevLett.129.121102} {\bibfield  {journal} {\bibinfo
  {journal} {Phys. Rev. Lett.}\ }\textbf {\bibinfo {volume} {129}},\ \bibinfo
  {pages} {121102} (\bibinfo {year} {2022})},\ \Eprint
  {http://arxiv.org/abs/2209.15487} {arXiv:2209.15487 [gr-qc]} \BibitemShut
  {NoStop}%
\bibitem [{\citenamefont {Afach}\ \emph {et~al.}(2021)\citenamefont {Afach}
  \emph {et~al.}}]{Afach:2021pfd}%
  \BibitemOpen
  \bibfield  {author} {\bibinfo {author} {\bibfnamefont {S.}~\bibnamefont
  {Afach}} \emph {et~al.},\ }\href {\doibase 10.1038/s41567-021-01393-y}
  {\bibfield  {journal} {\bibinfo  {journal} {Nature Phys.}\ }\textbf {\bibinfo
  {volume} {17}},\ \bibinfo {pages} {1396} (\bibinfo {year} {2021})},\ \Eprint
  {http://arxiv.org/abs/2102.13379} {arXiv:2102.13379 [astro-ph.CO]}
  \BibitemShut {NoStop}%
\bibitem [{\citenamefont {Arvanitaki}\ \emph {et~al.}(2015)\citenamefont
  {Arvanitaki}, \citenamefont {Huang},\ and\ \citenamefont
  {Van~Tilburg}}]{Arvanitaki:2014faa}%
  \BibitemOpen
  \bibfield  {author} {\bibinfo {author} {\bibfnamefont {A.}~\bibnamefont
  {Arvanitaki}}, \bibinfo {author} {\bibfnamefont {J.}~\bibnamefont {Huang}}, \
  and\ \bibinfo {author} {\bibfnamefont {K.}~\bibnamefont {Van~Tilburg}},\
  }\href {\doibase 10.1103/PhysRevD.91.015015} {\bibfield  {journal} {\bibinfo
  {journal} {Phys. Rev. D}\ }\textbf {\bibinfo {volume} {91}},\ \bibinfo
  {pages} {015015} (\bibinfo {year} {2015})},\ \Eprint
  {http://arxiv.org/abs/1405.2925} {arXiv:1405.2925 [hep-ph]} \BibitemShut
  {NoStop}%
\bibitem [{\citenamefont {Kennedy}\ \emph {et~al.}(2020)\citenamefont
  {Kennedy}, \citenamefont {Oelker}, \citenamefont {Robinson}, \citenamefont
  {Bothwell}, \citenamefont {Kedar}, \citenamefont {Milner}, \citenamefont
  {Marti}, \citenamefont {Derevianko},\ and\ \citenamefont
  {Ye}}]{Kennedy:2020bac}%
  \BibitemOpen
  \bibfield  {author} {\bibinfo {author} {\bibfnamefont {C.~J.}\ \bibnamefont
  {Kennedy}}, \bibinfo {author} {\bibfnamefont {E.}~\bibnamefont {Oelker}},
  \bibinfo {author} {\bibfnamefont {J.~M.}\ \bibnamefont {Robinson}}, \bibinfo
  {author} {\bibfnamefont {T.}~\bibnamefont {Bothwell}}, \bibinfo {author}
  {\bibfnamefont {D.}~\bibnamefont {Kedar}}, \bibinfo {author} {\bibfnamefont
  {W.~R.}\ \bibnamefont {Milner}}, \bibinfo {author} {\bibfnamefont {G.~E.}\
  \bibnamefont {Marti}}, \bibinfo {author} {\bibfnamefont {A.}~\bibnamefont
  {Derevianko}}, \ and\ \bibinfo {author} {\bibfnamefont {J.}~\bibnamefont
  {Ye}},\ }\href {\doibase 10.1103/PhysRevLett.125.201302} {\bibfield
  {journal} {\bibinfo  {journal} {Phys. Rev. Lett.}\ }\textbf {\bibinfo
  {volume} {125}},\ \bibinfo {pages} {201302} (\bibinfo {year} {2020})},\
  \Eprint {http://arxiv.org/abs/2008.08773} {arXiv:2008.08773
  [physics.atom-ph]} \BibitemShut {NoStop}%
\bibitem [{\citenamefont {Damour}\ and\ \citenamefont
  {Donoghue}(2010)}]{Damour:2010rm}%
  \BibitemOpen
  \bibfield  {author} {\bibinfo {author} {\bibfnamefont {T.}~\bibnamefont
  {Damour}}\ and\ \bibinfo {author} {\bibfnamefont {J.~F.}\ \bibnamefont
  {Donoghue}},\ }\href {\doibase 10.1088/0264-9381/27/20/202001} {\bibfield
  {journal} {\bibinfo  {journal} {Class. Quant. Grav.}\ }\textbf {\bibinfo
  {volume} {27}},\ \bibinfo {pages} {202001} (\bibinfo {year} {2010})},\
  \Eprint {http://arxiv.org/abs/1007.2790} {arXiv:1007.2790 [gr-qc]}
  \BibitemShut {NoStop}%
\bibitem [{\citenamefont {Damour}\ and\ \citenamefont
  {Polyakov}(1994)}]{Damour:1994zq}%
  \BibitemOpen
  \bibfield  {author} {\bibinfo {author} {\bibfnamefont {T.}~\bibnamefont
  {Damour}}\ and\ \bibinfo {author} {\bibfnamefont {A.~M.}\ \bibnamefont
  {Polyakov}},\ }\href {\doibase 10.1016/0550-3213(94)90143-0} {\bibfield
  {journal} {\bibinfo  {journal} {Nucl. Phys. B}\ }\textbf {\bibinfo {volume}
  {423}},\ \bibinfo {pages} {532} (\bibinfo {year} {1994})},\ \Eprint
  {http://arxiv.org/abs/hep-th/9401069} {arXiv:hep-th/9401069} \BibitemShut
  {NoStop}%
\bibitem [{\citenamefont {Brax}\ and\ \citenamefont
  {Martin}(2007)}]{Brax:2006dc}%
  \BibitemOpen
  \bibfield  {author} {\bibinfo {author} {\bibfnamefont {P.}~\bibnamefont
  {Brax}}\ and\ \bibinfo {author} {\bibfnamefont {J.}~\bibnamefont {Martin}},\
  }\href {\doibase 10.1103/PhysRevD.75.083507} {\bibfield  {journal} {\bibinfo
  {journal} {Phys. Rev. D}\ }\textbf {\bibinfo {volume} {75}},\ \bibinfo
  {pages} {083507} (\bibinfo {year} {2007})},\ \Eprint
  {http://arxiv.org/abs/hep-th/0605228} {arXiv:hep-th/0605228} \BibitemShut
  {NoStop}%
\end{thebibliography}%

\end{document}